\documentclass[reprint,superscriptaddress,amsmath,amssymb,aps,prr,floatfix]{revtex4-2}

\usepackage{graphicx}
\usepackage{physics}
\usepackage{amsmath, amssymb}
\usepackage{amsthm}
\usepackage{dsfont}
\usepackage{subfig}
\usepackage{tabularx}
\usepackage{units}
\usepackage[dvipsnames]{xcolor}

\definecolor{myblue}{named}{NavyBlue}
\definecolor{mygreen}{RGB}{13,255,13}
\usepackage[colorlinks=true,citecolor=myblue,linkcolor=myblue,urlcolor=myblue]{hyperref}
\usepackage[english, capitalise]{cleveref}

\newcommand{\R}{\mathbb{R}}
\newcommand{\C}{\mathbb{C}}
\newcommand{\N}{\mathbb{N}}
\newcommand{\Hil}{\mathcal{H}}

\newcommand{\1}{\mathds{1}}
\DeclareMathOperator{\diag}{diag}

\newcommand{\Nin}{N_{\mathrm{in}}}
\newcommand{\Nout}{N_{\mathrm{out}}}
\newcommand{\kmax}{k_{\text{max}}}

\newcommand{\mbf}[1]{\mathbf{#1}}

\theoremstyle{definition}
\newtheorem{defn}{Definition}

\theoremstyle{plain}
\newtheorem{thrm}[defn]{Theorem}

\theoremstyle{plain}
\newtheorem{lem}[defn]{Lemma}

\theoremstyle{plain}
\newtheorem{cor}[defn]{Corollary}

\theoremstyle{definition}
\newtheorem{prot}{Protocol}

\begin{document}
	
	\title{Improved Decoy-state and Flag-state Squashing Methods}
	
	\author{Lars Kamin}
	\email{lars.kamin@outlook.com}
	\affiliation{Institute for Quantum Computing and Department of Physics and Astronomy, University of Waterloo, Waterloo, Ontario N2L 3G1, Canada}

	\author{Norbert L{\"u}tkenhaus}
	\email{lutkenhaus.office@uwaterloo.ca}
	\affiliation{Institute for Quantum Computing and Department of Physics and Astronomy, University of Waterloo, Waterloo, Ontario N2L 3G1, Canada}
	
	\date{\today}
	
	\begin{abstract}
		In this work, we present an improved analysis for decoy-state methods, enhancing both achievable key rates and recovering analytical results for the single intensity scenario \cite{Lutkenhaus2000Phys.Rev.A}. Our primary focus is improving the shortcomings observed in current decoy-state methods, particularly recovering results when employing no decoy intensities. Our methods enable the continuous interpolation across varying numbers of intensity settings. Additionally, we extend decoy-state techniques to encompass scenarios where intensities vary depending on the signal state, thereby relaxing the constraints on experimental implementations. Our findings demonstrate that a minimum of two intensities are sufficient for high asymptotic secret key rates, thereby further softening experimental requirements.
		
		Additionally, we address inherent imperfections within detection setups like imperfect beamsplitters. We derive provable secure lower bounds on the subspace population estimation, which is required for certain squashing methods such as the flag-state squasher \cite{Zhang2021Phys.Rev.Res.}. These analytical bounds allow us to encompass arbitrary passive linear optical setups, and together with intensities varying with each signal state, lets us include a broad class of experimental setups.
	\end{abstract}
	
	\maketitle
	
\section{Introduction}
Weak coherent pulse (WCP) sources are the most commonly used signal sources for quantum key distribution (QKD). Frequently, WCP sources are used that emit fully phase-randomised states and we will work in that scenario. However, WCP sources are susceptible to the photon-number-splitting (PNS) attack \cite{Brassard2000Phys.Rev.Lett.}. Due to the PNS attack an adversary could gain full information of any multi-photon state sent without revealing themselves. Therefore, several techniques have been developed to mitigate this issue.

If one keeps the experimental setup the same and simply accepts the possibility of a PNS attack, security can still be proven as in Ref.~\cite{Lutkenhaus2000Phys.Rev.A}. However, the secret key rates are much lower than those of a perfect single-photon source.

For this reason, the alternative of decoy-state protocols \cite{Hwang2002Phys.Rev.Lett.,Ma2005Phys.Rev.A,Wang2005Phys.Rev.Lett.} have been developed. Decoy-state methods involve Alice sending multiple different intensities to perform partial channel tomography and identifying the single-photon contribution to the detection events. Due to the structure of the states emitted by fully phase randomized WCP sources, an adversary (Eve) cannot distinguish the signal intensity chosen for a particular signal by just looking at the photon number of that signal. Thus, Eve's optimal attack is attacking based on the photon number and the additional observations from decoy intensities allow one to bound her influence on the single-photon signals.

The first decoy-state methods used analytic techniques to bound the single-photon yield as well as the error rate. Later, the decoy analysis was reformulated as a linear program in Refs.~\cite{Rice2009, Wang2022Phys.Rev.Res.} which is typically solved numerically, and tighter than the analytical bounds. Furthermore, the techniques of \cite{Wang2022Phys.Rev.Res., Rice2009} could incorporate more fine-grained data than simple error rates.

On the other hand, the methods of Ref.~\cite{Lutkenhaus2000Phys.Rev.A} chose a different route to bound the influence on the single-photon signals. Since multi-photon signals are fully insecure, they were tagged and thus, announced to the adversary. Then, minimizing the possible attacks an adversary could perform, lead to an analytic key rate for a WCP source without any decoy states.

However, an unresolved problem of the decoy-state analysis was, that it could not recover the results from Ref.~\cite{Lutkenhaus2000Phys.Rev.A} and therefore, was not tight in both the resulting key rates and the number of intensities used. Moreover, even the tighter decoy-state methods presented in \cite{Rusca2018Appl.Phys.Lett.}, which use additional information on the vacuum error rate, do not recover the key rate of a single weak coherent pulse completely, although they are already an important step towards it.

In this work we will present a refined decoy-state analysis which fully recovers the no-decoy WCP rate of Ref. \cite{Lutkenhaus2000Phys.Rev.A}. Moreover, it remained generally unknown how many decoy intensities are actually necessary to reach a key rate close to the infinite decoy limit with a fully phase-randomized WCP source or equivalently a simple beam-splitting attack. So far, two decoy intensity settings proved to be good in practice. We will show that already two intensities will suffice for both the BB84 and six-state protocol. Both partial results of this work have already been presented as a poster at QCrypt22 \cite{Poster}.

We only consider the asymptotic regime, but our methods can be easily extended to the finite-size regime. The main idea of our approach is stated in \cref{sec:Improved Decoy Methods}, where we reformulate the decoy analysis as an SDP to use the full information available to us by all observations. Instead of generalizing the standard decoy formulated as linear programs one could also use the generalised decoy methods presented in \cite{Nahar2023} and simplify it using the block-diagonal structure of the signal states.

Additionally, to incorporate the experimentally common problem of having unequal intensities depending on the state sent, we extend our approach to apply to this case as well. For this purpose, we derive a new key rate formula in \cref{sec:Extension of Asymptotic Key Rate with Differing Intensities}, which will still be valid if the intensities differ between each signal sent by Alice.

Any numerical security proof for optical implementations requires a so-called squashing map, which maps the infinite dimensional Hilbert space of the optical modes to a finite-dimensional  space. One such method is the flag-state squasher \cite{Zhang2021Phys.Rev.Res.}, which can be applied to any discrete variable protocol. It maps the infinite-dimensional Hilbert space to a finite-dimensional one by leaving a subspace invariant and replacing all parts of the incoming state outside that subspace with classical flags. Thereby, the flag-state squasher effectively announces any measurement result outside the subspace. Hence, for QKD applications it is required to bound the weight inside the subspace to obtain any positive key rate. Therefore, we extend previous results on the flag-state squasher to arbitrary passive linear optical setups, which is summarized in \cref{Thrm:Lower bound flag state perfect} and \cref{Thrm:Lower bound flag state lossy}. In addition, we present a procedure how the necessary quantities can be experimentally measured and then used for a detector characterization in \cref{App:Detector Characterization for Flag-State Squasher}.

The structure of this paper is the following, in \cref{sec:Protocol description} and \cref{sec:Asymptotic Key Rate with Decoy States} we lay out the structure of a QKD protocol and restate the known results for the secret key rate using decoy-state methods from \cite{Li2020Phys.Rev.Research} and summarize previous decoy methods formulated as linear programs. Next, in \cref{sec:Improved Decoy Methods} we generalize those techniques and introduce our improved decoy methods, where we also show how our methods recover certain imposed assumptions on vacuum error rates. We conclude this initial part in \cref{sec:Example 1: BB84} with the example of the BB84 protocol and show the resulting key rates recover the no-decoy rates from \cite{Lutkenhaus2000Phys.Rev.A}.

Next, in \cref{sec:Extension of Asymptotic Key Rate with Differing Intensities} and \cref{sec:Flag-state squasher}, we focus on imperfections in both the state generation and detection setups, i.e. we derive the asymptotic key rate formula for intensities differing with each signal and construct the flag-state squasher for arbitrary passive linear optical setups. To conclude this part, we present in \cref{sec:Example 2: Biased passive 6-state} the six-state protocol with biased basis choices and differing intensities as examples. Finally, in \cref{sec:Conclusion} we discuss and summarize the results.

\section{Protocol description}\label{sec:Protocol description}
For completeness, we describe the steps of a generic P\&M protocol using decoy states in \cref{Prot:PM Protocol}. We assume that a classical authenticated channel is available as a resource to this QKD protocol.

\begin{prot}[Generic Decoy-State Prepare-and-Measure Protocol]\label{Prot:PM Protocol}
\textbf{Protocol steps:}
\begin{enumerate}
	\item \textbf{State preparation and transmission:} In each round Alice decides with probability \(p(\text{gen})\) whether she sends a test or key generation round. In the case of a key generation round, Alice prepares one of \(d_A\) states \(\{\ket{s_i}\}_{i=1\dots d_A}\) with intensity \(\mu_s\). For test rounds Alice selects the intensity \(\mu_i\) with probability \(p(\mu_i |\text{test})\). She stores a label for her signal state in register \(X_i\). Depending on her choices Alice computes her announcement and stores it in register \(C_A\). Finally, Alice sends the signal state to Bob via a quantum channel.
	\item \textbf{Measurements:} Bob measures his received states using a Positive Operator Valued Measure (POVM) with POVM elements \(\{F_k^B\}_{k=1 \dots d_B}\), and stores his measurement outcomes in a register \(Y\) with alphabet \(\mathcal{Y}\). Furthermore, he computes classical data for public announcement and stores it in \(C_B\).
\end{enumerate}

Alice and Bob repeat steps 1. and 2. \(N\) times
\begin{enumerate}
	\setcounter{enumi}{2}
	\item \textbf{Public announcement and sifting:} Alice and Bob have a public discussion (which can be multi-round) and announce their public data contained in \(C\), computed from \(C_AC_B\). Then, they sift their secret data based on their public communication.
	\item \textbf{Acceptance test (parameter estimation):} Alice and Bob perform statistical tests on the randomly chosen subset of the data, based on Alice's decision in step 1. They test if \(F^{\text{obs}} \in \mathcal{Q} \), where \(\mathcal{Q}\) is the predetermined acceptance set. Depending on their results Alice and Bob continue or abort the protocol.
	\item \textbf{Key map:} Alice (or Bob) performs the key map \(g:\mathcal{XC} \rightarrow \mathcal{R}\), where \(R\) is the key register with alphabet \(\mathcal{R}\).
	\item \textbf{Error correction and verification:} Using the authenticated channel, Alice and Bob communicate to reach agreeing versions of the key register \(R\) for both parties. In the process, they send some error correction information. After the error correction step, they send a hash to verify the success of the correction procedure. If error correction was successful, this check succeeds except with small probability \(\varepsilon_{\text{EV}}\). In total they reveal \(\delta_{\text{leak}}\) bits per channel use.
	\item \textbf{Privacy Amplification:} Alice and Bob randomly choose a universal hash function and apply it their raw keys, producing their final keys of fixed secure length \(l\).
\end{enumerate}
\end{prot}

\section{Asymptotic Key Rate with Decoy States}\label{sec:Asymptotic Key Rate with Decoy States}

First, let us reintroduce the key rate formula from \cite{Wang2022Phys.Rev.Res.} for decoy-state methods. Let us assume in the key generation rounds Alice is sending WCP signals with the signal intensity \(\mu_{s}\). Then, using the source replacement scheme and including a shield system \(A_s\), one can write the signal state as
\begin{equation}\label{eq:Source Replacement}
	\ket{\xi}_{AA_sA'} = \sum_{x,n} \sqrt{p_x p_{\mu_s}(n)} \ket{x}_A \ket{n}_{A_s} \ket{\psi^{(x,n)}}_{A'},
\end{equation}
where \(A\) and \(A'\) represent Alice's systems and the system to be sent to Bob, respectively. As shown in \cite{Wang2022Phys.Rev.Res.}, due to the phase randomization, the state sent is a mixture of Fock states and the photon number is known to Eve due to possible QND measurements. Thus, the final state between Alice (\(A\)) and Bob (\(B\)) is block diagonal in the photon number and given by
\begin{equation}
	\rho_{AA_sB} = \sum_{n} p_n \dyad{n}_{A_s} \otimes \rho_{AB}^{(n)}.
\end{equation}
Furthermore, let \(\{\Gamma_{x,y}\}_{xy} = \{\dyad{x} \otimes F_y^B \}_{xy}\) be the POVM elements acting on systems \(A\) and \(B\), corresponding Alice's and Bob's observations \(\{\gamma_{x,y}\} \). 
Hence, following \cite[App. B]{Li2020Phys.Rev.Research} the key rate is 
\begin{equation}\label{eq:Key Rate decoy}
	R \geq \sum_{n=0}^{\infty} p_n \min_{\rho_{AB}^{(n)} \in S_n} f\left(\rho_{AB}^{(n)}\right) - p_{\text{pass}} \delta_{\text{leak}},
\end{equation}
where \(f(\rho) := D\left(\mathcal{G}\left(\rho \right) || \mathcal{G}\left(\mathcal{Z}\left(\rho \right) \right) \right)\) and \(D(.||.)\) is the relative entropy. The optimizations run over the feasible sets \(S_n\) conditioned on \(n\) photons sent by Alice, which are constructed from the POVMs \(\{\Gamma_{x,y}\}_{xy}\) and the observations \(\{\gamma_{x,y}\}_{xy}\) and will be discussed in more detail below. To evaluate this optimization for given \(S_n\) and therefore the asymptotic key rate, we use the methods from \cite{Winick2018Quantum}.

If for each \(n\) the sets \(S_n\) would allow for arbitrary states albeit the observations, the key rate would obviously become zero. Hence, the most important step is finding a tight description of the sets \(S_n\).

As an first example of this task, we introduce previous decoy methods \cite{Wang2022Phys.Rev.Res.}, which have been phrased as linear programs. Define
\begin{equation}
	\gamma_{y|x}^{\mu} := \Pr(y|x,\mu)
\end{equation}
as the probability of observing outcome \(y\) given state \(x\) with intensity \(\mu\) was sent by Alice. Then, define the \(n\)-photon yields
\begin{equation}
	Y_n^{x,y} := \Pr(y|x,n),
\end{equation}
as the conditional probability of observing outcome \(y\) given Alice sent state \(x\) with \(n\) photons.

Now we can describe the feasible sets \(S_n\) in \cref{eq:Key Rate decoy} in more detail. If we had perfect knowledge of the \(n\)-photon yields the feasible set \(S_n\) for each photon number \(n\) would be given by 
\begin{equation}
	\begin{aligned}
		S_n = \{ &\rho_{AB}^{(n)} \in \mathcal{D}(\Hil_A \otimes \Hil_B)| \\ 
		&\hspace{20pt} \Tr\left[ \Gamma_{x,y} \rho_{AB}^{(n)} \right] = p(x) Y^{x,y}_n\},
	\end{aligned}
\end{equation}
where additional constraints like the normalization have been left out for simplicity. Here, again the operators \(\{\Gamma_{x,y}\}_{xy}\) are Alice and Bob's POVM elements.

However, the \(n\)-photon yield is not directly observable, but we can bound it by observations by deriving independent upper and lower bounds from the observations as in \cite{Wang2022Phys.Rev.Res.}
\begin{equation}
	\begin{aligned}
		S_n &= \{ \rho_{AB}^{(n)} \in \mathcal{D}(\Hil_A \otimes \Hil_B) \;| \\ 
		&\Pr(x) Y^{x,y}_{n,L} \leq \Tr\left[ \Gamma_{x,y} \rho_{AB}^{(n)} \right] \leq \Pr(x) Y^{x,y}_{n,U},\\
		&\Tr_B(\rho_{AB}) = \sigma_A \}.
	\end{aligned}
\end{equation}
where \(\sigma_A\) characterizes the reduced state constraints due to the source replacement scheme used for P\&M protocols for example see \cite[Sec. IV C]{Li2020Phys.Rev.Research}. The upper and lower bounds \(Y^{x,y}_{n,U/L}\) will be derived in the remaining part of this subsection. However, it is noteworthy at this point, that the bounds \(Y^{x,y}_{n,U/L}\) we present are independent from each other for each combination of observations \(x,y\), which is a simplification that will result in some looseness.

Thus, given the shape of the sets \(\tilde{S}_n\) and in order to calculate the asymptotic key rate, we are left with the problem of finding the upper and lower bounds of the \(n\) photon yields. Typically, only the single-photon yields are considered, as due to the PNS attack all multi-photon signals are insecure and do not give a positive contribution to the key rate, and vacuum signals only contribute on the order of dark counts.

The \(n\)-photon yields and the observations \(\gamma_{y|x}^{\mu}\) are related by
\begin{equation}
	\gamma_{y|x}^{\mu} = \sum_{n=0}^{\infty} p_{\mu}(n) Y_n^{x,y}.
\end{equation}
Since one cannot solve a system of equations with infinitely many variables, one instead uses the following upper and lower bounds given a photon number cut-off \(N_{\text{ph}}\),
\begin{align}
	\gamma_{y|x}^{\mu} &\geq \sum_{ n \leq N_{\mathrm{ph}}} p_{\mu}(n) Y_n^{x,y}, \\
	\gamma_{y|x}^{\mu} &\leq \sum_{ n \leq N_{\mathrm{ph}}} p_{\mu}(n) Y_n^{x,y} + (1-p_{tot}(\mu)),
\end{align}
where \(p_{\mathrm{tot}} = 1-\sum_{n\leq N_{\mathrm{ph}}} p_{\mu} (n) \) was introduced.

Hence, the lower bound on the single photon yield of observation \(i,j\) can be found by the following minimization problem
\begin{equation}\label{eq:LP decoy}
	\begin{aligned}
		Y_{n,L}^{i,j}:=&\min_{\mbf{Y}_m}\;   Y^{i,j}_n\\
		\textrm{s.t.}\; &  \gamma^{\mu}_{j|i} \leq \sum_{m\leq N_{\mathrm{ph}}} p_{\mu}(m)\; Y_m^{i,j} + (1-p_{tot}(\mu)), \\
		&\gamma^{\mu}_{j|i} \geq \sum_{m\leq N_{\mathrm{ph}}} p_{\mu}(m)\; Y_m^{i,j}, \\
		&\forall \mu \in \{\mu_1, \mu_2, \dots\}, \\
		&0 \leq Y^{i,j}_m \leq 1 \; \forall m \in \N_0,
	\end{aligned}    
\end{equation}
and a maximization problem with the same constraints has to be solved for the upper bound. This simple independent linear program achieves tight bounds for the sets \(\tilde{S}_n\) with three and more intensities. Therefore, one also achieves tight asymptotic key rates in those cases.

However, this formulation does not produce positive key rate with one single intensity, i.e. does not recover the no-decoy WCP rate of \cite{Lutkenhaus2000Phys.Rev.A} as already mentioned in the introduction. Again, one would expect that there exists a method connecting both approaches and reducing to the no-decoy WCP rate if only a single intensity setting is used in the decoy-state methods. In the next \cref{sec:Improved Decoy Methods}, we show a method which recovers this limit.

The reason for the shortcoming of the linear program \cref{eq:LP decoy} is that each optimization problem takes only the observations for different intensities of that particular observable into account. Thus, one ignores constraints between different observations, for instance in the loss-only scenario for unbiased BB84 one expects the sum of observations of different bases to be equal. Hence, we do not use all information available to us, which leads to no key rate for the no-decoy setting.

In contrast, the no-decoy approach of \cite{Lutkenhaus2000Phys.Rev.A} uses the vacuum error rate and the detection probability jointly, thereby creating a cross connection between different observables. Hence, the resulting bounds on the yields are not independent anymore in contrast to \cref{eq:LP decoy}.

\section{Improved Decoy Methods}\label{sec:Improved Decoy Methods}
Now, we introduce our improved decoy methods. To reiterate, we rely on fully phase-randomised coherent states. To incorporate states which are not fully phase-randomized, the techniques of \cite{Nahar2023} can be used.

Since we consider fully phase-randomized coherent states, the signal states are block diagonal in the photon number as seen in \cref{eq:State Alice Bob}. Thus, Eve's channel can be recast as a direct sum acting on each block separately
\begin{equation}
	\mathcal{E} = \bigoplus_{n=0}^{\infty} \mathcal{E}_n,
\end{equation}
where each \(\mathcal{E}_n : \Hil_{A'_n} \rightarrow \Hil_{B_M} \). This decomposition of Eve's channel holds because Eve could always apply a QND measurement to detect the photon number and then adjust her attack based on the photon number \cite{Li2020Phys.Rev.Research}. Here, \(\Hil_{B_M}\) is Bob's infinite dimensional state space.

In order to treat the infinite-dimensional full space \(\Hil_{B_M}\) numerically, a squashing map is needed. A squashing map acts on \(\Hil_{B_M}\) and maps it to a finite-dimensional space \(\Hil_B\), which can be a qubit space. Mathematically, a squashing map \(\Lambda\) acts as
\begin{equation}
	\Lambda: \mathcal{H}_{B_M} \rightarrow \mathcal{H}_{B},
\end{equation}
where now \( \mathcal{H}_{B}\) is \emph{finite-dimensional}. To keep the observations invariant under the squashing map, it is required that
\begin{equation}
	\Tr[\rho_{\text{in}} F_{B_M}^{(k)} ] = \Tr[\Lambda(\rho_{\text{in}}) F_B^{(k)} ],
\end{equation}
for all input states \(\rho_{\text{in}}\) and POVM elements \(k\). For optical implementations of QKD protocols using threshold detectors, the squashing maps presented in \cite{Gittsovich2014Phys.Rev.A, Beaudry2008Phys.Rev.Lett.} can be applied under certain conditions and then, \( \mathcal{H}_{B}\) simplifies to a qubit space.

However, in \cref{sec:Example 2: Biased passive 6-state}, we will present an example where these squashing maps cannot be applied and different squashing techniques are required. Due to the block-diagonal structure of Eve's channel, the full channel (consisting of Eve's attack and the squashing map) has block-diagonal form 
\begin{equation}
	\Phi = \bigoplus_{n=0}^{\infty} \Phi_n = \bigoplus_{n=0}^{\infty} \Lambda \circ \mathcal{E}_n: \mathcal{H}_{A'} \rightarrow \mathcal{H}_{B}.
\end{equation}
By including the squashing map in the full channel, we give it into the hands of the adversary, Eve, and are only overestimating their power.

However, this enables us to use the Choi isomorphism for each channel \(\Phi_n\) separately to represent its action by the Choi matrix \(J_n\). Hence, the states in each set \(S_n\) must be generated by the same channel \(J_n\) acting on the states \(\ket{\psi^{(x,n)}}_{A'}\) of \cref{eq:Source Replacement} sent by Alice. 

This allows us to rewrite the \(n\)-photon yields for all \(x,y\) as
\begin{align}
	Y_n^{x,y} &= \Tr\left[F_y^B \Phi_n\left(\rho_x^{(n)}\right) \right] \\
	&=  \Tr\left[J_n \left(F_y^B \otimes \left(\rho_x^{(n)}\right)^T \right)\right],
\end{align}
where \(F_y^B\) are Bob's POVM elements and \(\rho_x^{(n)} =\ketbra{\psi^{(x,n)}}\) is the density matrix of the state sent by Alice.

Connecting the constraints on the yields with each other via the attack channel is the most crucial difference of our methods compared to previous ones. The channel \(\Phi_n\) has to be the same for all observations. Hence, we can include all information about the channel in the optimization problem and any correlations between observations are incorporated automatically. The final minimization/maximization problem for \(n\)-photon yield can be written as
\begin{equation}\label{eq:SDP decoy}
	\begin{aligned}
		Y_{n,L}^{i,j}:=&\min_{\mbf{Y}_m, J_m}\;  Y^{i,j}_{n}\\
		\textrm{s.t.}\; &  \gamma^{\mu}_{y|x} \leq \sum_{m\leq N_{\mathrm{ph}}} p_{\mu}(m)\; Y_m^{x,y} +(1-p_{tot}), \\
		&\gamma^{\mu}_{y|x} \geq \sum_{m\leq N_{\mathrm{ph}}} p_{\mu}(m)\; Y_m^{x,y}, \\
		&Y_m^{x,y} = \Tr\left[J_m \left(F_y^B \otimes \left(\rho_x^{(m)}\right)^T \right)\right], \\
		&0 \leq Y^{x,y}_m \leq 1, \\
		&J_m \succeq 0, \; \Tr_B\left[J_m \right] = \1_{A'}, \\ 
		&\forall \mu \in \{\mu_1, \mu_2, \dots\}, \; \forall x,y, \\
		&\forall \;  0\leq m \leq N_{\mathrm{ph}},
	\end{aligned}    
\end{equation}
which is convex in the Choi matrices \(J_n\). For the maximization problem, the minimum is simply replaced with a maximum. 

The additional partial trace and positivity constraints on the Choi matrix result from the fact that we require the full channel \(\Phi_n\) to be a CPTP map. These additional requirements on the channel are another difference to previous methods.

Moreover, note that in \cref{eq:SDP decoy} the constraints run over all observations \(x,y\) in contrast to \cref{eq:LP decoy}. In \cref{eq:LP decoy} only the observation corresponding to the detection event \(i,j\) and the resulting constraints of that particular entry were considered. Already from this it becomes apparent that we use much more information about the channel in each minimization.

\section{Example 1: WCP Decoy BB84}\label{sec:Example 1: BB84}

Again, for all examples presented, we use the methods of \cite{Winick2018Quantum} to calculate the asymptotic key rate. More details how to include decoy-state methods into this framework, in particular how the BB84 protocol is included can be found in \cite{Wang2022Phys.Rev.Res.}.

We start with simple unbiased active BB84 to illustrate the idea of our approach. We consider the following parameters for the BB84 protocol, \(p_x = p_z =\frac{1}{2}\) for the basis choices and the Shannon limit of \(f_{\text{EC}} = 1\) as the error correction efficiency. The choice of intensity settings are always optimized for each data point. For all of our results in this section we  assume a loss parameter of \(\unit[0.2]{dB/km}\).

For this optical setup, we use the squashing map from \cite{Gittsovich2014Phys.Rev.A,Beaudry2008Phys.Rev.Lett.}, which is block diagonal in the photon number as well and Bob's POVM elements \emph{after squashing} are the usual qubit POVM elements i.e.
\begin{align}
	\{F_y^B \} = \{ \dyad{H}, \dyad{V}, \dyad{+}, \dyad{-} , \dyad{\mathrm{vac}}\}.
\end{align}

We simplify \cref{eq:SDP decoy} and only consider Choi matrices for \(n=0\) and \(n=1\) photons. This captures the essential action of the channel as we will see in the results and keeps it numerically feasible. Then the simplified minimization/maximization problem for \(n\)-photon yield becomes
\begin{equation}\label{eq:SDP BB84}
	\begin{aligned}
		Y_{n,L}^{i,j}:=&\min_{\mbf{Y}_m, J_0, J_1}\; Y^{i,j}_{n}\\
		\textrm{s.t.}\; &  \gamma^{\mu}_{y|x} \leq \sum_{m\leq N_{\mathrm{ph}}} p_{\mu}(m)\; Y_m^{x,y} +(1-p_{tot}), \\
		&\gamma^{\mu}_{y|x} \geq \sum_{m\leq N_{\mathrm{ph}}} p_{\mu}(m)\; Y_m^{x,y}, \\
		&Y_0^{x,y} = \Tr\left[J_0 \left(F_y^B \otimes \left(\rho_x^{(0)}\right)^T \right)\right], \\
		&Y_1^{x,y} = \Tr\left[J_1 \left(F_y^B \otimes \left(\rho_x^{(1)}\right)^T \right)\right], \\
		&0 \leq Y^{x,y}_m \leq 1,\; \forall \; 0\leq m \leq N_{\mathrm{ph}}, \\
		&J_0,\; J_1 \succeq 0, \Tr_B\left[J_0 \right] = 1, \; \Tr_B\left[J_1 \right] = \1_{A'}, \\
		&\forall \mu \in \{\mu_1, \mu_2, \dots\}, \; \forall x,y.
	\end{aligned}
\end{equation}
Before we present the results, let us consider this optimization problem in a little more detail and compare it to previous techniques. 

For the analytical approach in \cite{Lutkenhaus2000Phys.Rev.A} it was an important observation that the error rate for vacuum states is $e_0 = \frac{1}{2} $. With our approach using the Choi representation we recover this property. For instance, consider the vacuum error rate in HV-basis. Since \(\rho_x^{(0)}\) is equal for all of Alice's state choices \(x\), thus one finds
\begin{align}
	Y_0^{xH} \equiv p_H, \quad Y_0^{xV} \equiv p_V \quad \forall x .
\end{align}
Therefore, the vacuum error rate in the \(HV\) basis amounts to
\begin{align}
	e_0^{HV} = \frac{Y_0^{HV} + Y_0^{VH}}{Y_0^{HH} + Y_0^{HV} + Y_0^{VH} + Y_0^{VV}} = \frac{1}{2}.
\end{align}
The same applies for the DA-basis. In \cite{Rusca2018Appl.Phys.Lett.}, $e_0 = \frac{1}{2} $ was also used, and it was shown to result in improvements in the finite-size regime. However, the constraint \(e_0 = \frac{1}{2}\) had to be imposed on the problem by hand. In contrast, our methods naturally include this and any other constraint resulting from correlations between the observations.

\begin{figure}
	\centering
	\includegraphics[width=\linewidth]{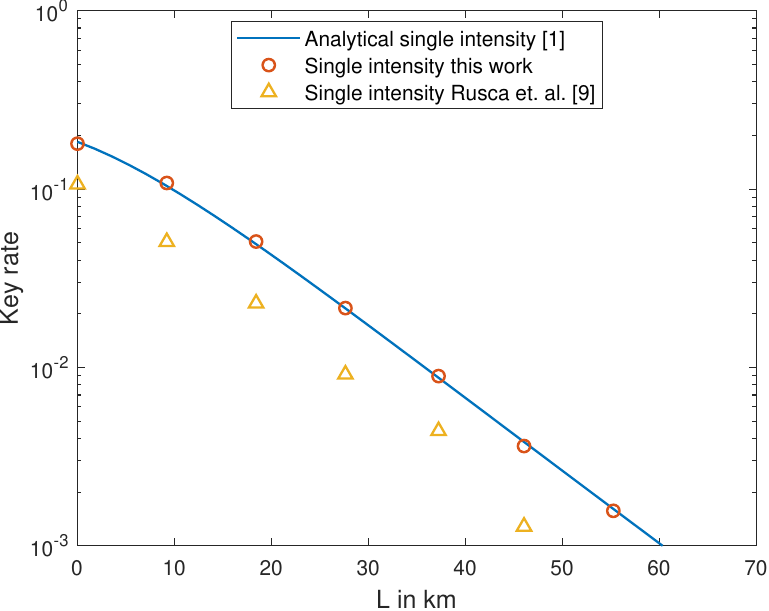}
	\caption{Secret key rate of the BB84 protocol over distance of our numerical results without decoy states (red circles) compared with numerical results from \cite{Rusca2018Appl.Phys.Lett.} (yellow triangles) and analytical results from \cite{Lutkenhaus2000Phys.Rev.A} (blue line).}
	\label{fig:BB84 single intensity}
\end{figure}

\begin{figure}
	\centering
	\includegraphics[width=\linewidth]{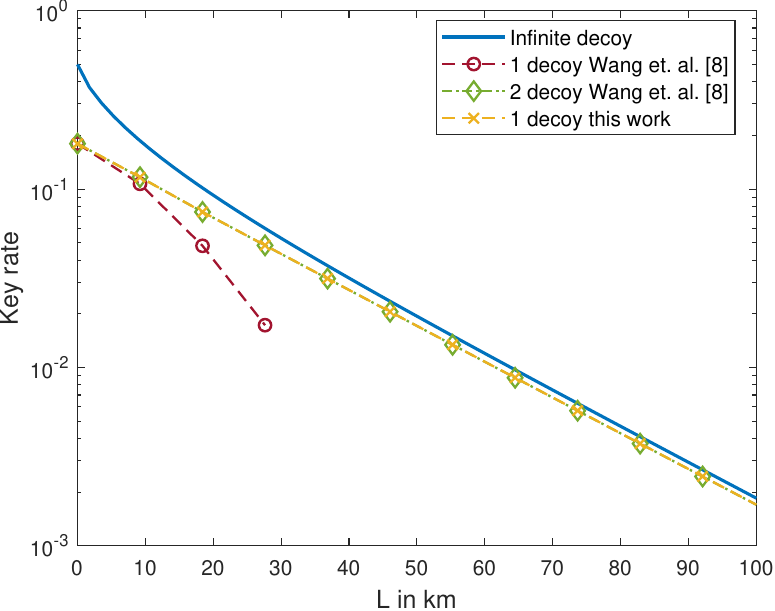}
	\caption{Comparison of secret key rates of the BB84 protocol over the distance of our numerical results for two intensities (yellow crosses) with numerical results from \cite{Wang2022Phys.Rev.Res.} using two (red circles) and three intensities (green diamonds).}
	\label{fig:BB84 comparison decoy methods}
\end{figure}

Now, let us turn to our numerical results. The resulting key rates for our decoy methods can be seen in \cref{fig:BB84 single intensity} and \cref{fig:BB84 comparison decoy methods} plotted against the distance in \(\unit{km}\). First, in \cref{fig:BB84 single intensity} the analytical result from \cite{Lutkenhaus2000Phys.Rev.A} (blue line) is compared with our numerical solution (red circles). As it can be clearly seen, we recover the analytical result. Moreover, we included the resulting key rate using the methods from \cite{Rusca2018Appl.Phys.Lett.} (yellow triangles), which do not recover the analytical result. This supports our claim that our approach of considering more information contained in the observations indeed yields a higher key rate.

Next, we focus on the comparison of different decoy methods and the required number of decoy intensities. First, in \cref{fig:BB84 comparison decoy methods} our decoy methods are compared with those from Wang et. al. \cite{Wang2022Phys.Rev.Res.}. We see a significant improvement, as we only need one decoy intensity to reach key rates equivalent to the previous approach using two decoy intensities. This showcases the improvements made since the methods of \cite{Wang2022Phys.Rev.Res.} achieve significantly lower key rates in the single-decoy scenario (red circles).

\begin{figure*}[ht]
	\centering
	\subfloat[Comparison between results for 1 decoy intensity using our improved decoy methods and two decoy intensities using previous methods from \cite{Wang2022Phys.Rev.Res.}. Here \(\text{Ratio} = \frac{\text{SKR}_{\text{1-dec}} - \text{SKR}_{\text{2-dec}}}{\text{SKR}_{\text{1-dec}}} \).]{\includegraphics[width=0.48\linewidth]{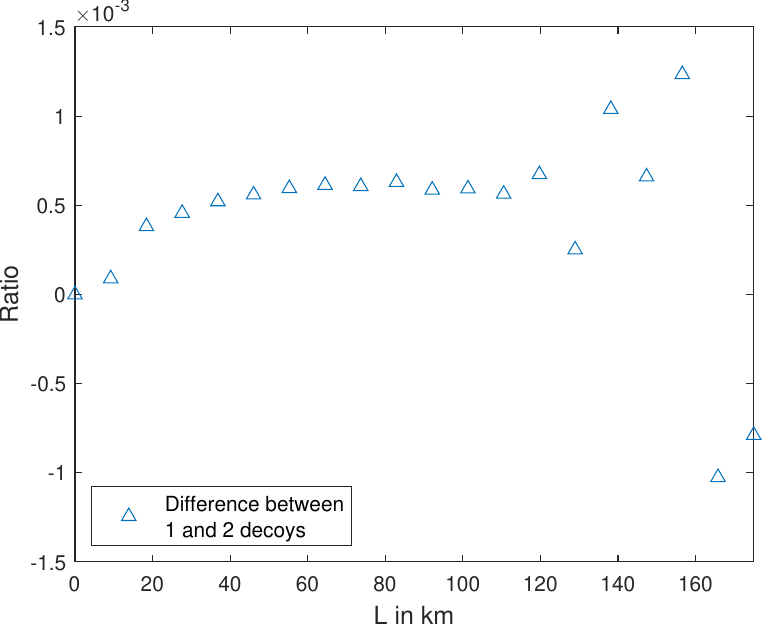}\label{fig:Difference BB84 2 impr vs 3 old}}
	\hfill
	\subfloat[Comparison between results for one and 2 decoy intensities using our improved decoy methods. Here \(\text{Ratio} = \frac{\text{SKR}_{\text{2-dec}} - \text{SKR}_{\text{1-dec}}}{\text{SKR}_{\text{2-dec}}} \).]{\includegraphics[width=0.48\linewidth]{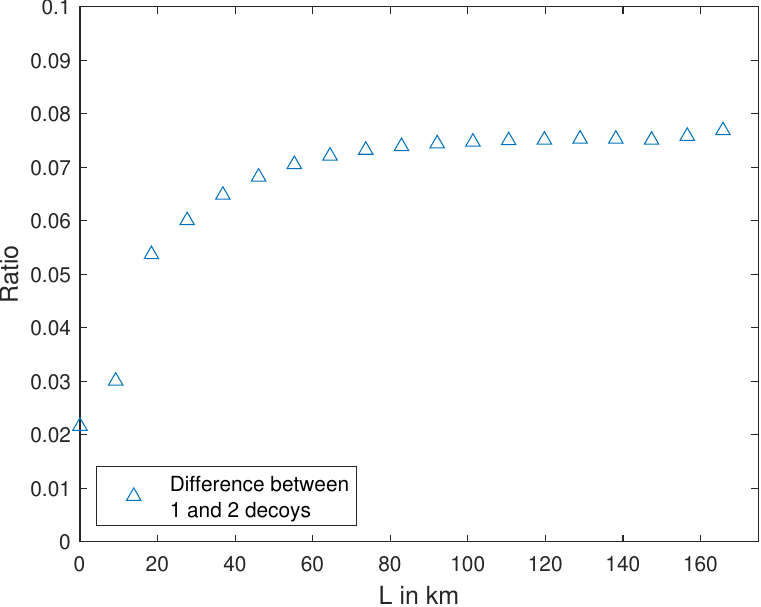}\label{fig:Difference BB84 2 impr vs 3 impr}
	}
	\caption{Comparison between the secret key rates of the BB84 protocol with one decoy intensity resulting from our improved analysis and two decoy intensities using both previous and our methods.}
	\label{fig:Difference BB84 2 vs 3}
\end{figure*}

In order to resolve the quantitative difference between our methods with a single decoy and those of \cite{Wang2022Phys.Rev.Res.} in more detail, we show the ratio of both results in \cref{fig:Difference BB84 2 impr vs 3 old}. One can see that a single decoy intensity can even yield a \(5\cdot 10^{-4} \%\) improvement in the key rate thanks to our improved methods. The different in terms of key rate may be insignificant, but using only one single decoy intensity significantly simplifies an experimental setup. Therefore, in order to maximize key rates and simplify the experimental setup, one should utilize our decoy methods.

At last, so far two decoy intensities were considered to be optimal, and despite the significant increase in key rate for a single decoy we have shown in \cref{fig:BB84 comparison decoy methods}, two decoy intensities could still be better. Therefore, in \cref{fig:Difference BB84 2 impr vs 3 impr}, we compare the difference between one and two decoy intensities both incorporating our improved decoy methods. Especially, for high losses two decoys produce a key rate that is \(7.5\%\) higher. 

\section{Extension of Asymptotic Key Rate with Differing Intensities}\label{sec:Extension of Asymptotic Key Rate with Differing Intensities}
Now, we turn our attention to errors in the preparation and detection of the signals and how to incorporate those. First, the previous results in \cite{Wang2022Phys.Rev.Res.,Li2020Phys.Rev.Research} did not allow for a decoy-state analysis if the intensities differed for each signal. We will derive an asymptotic key rate expression for this case, which is motivated by experimental realizations showing this feature. It recovers the original expression as a special case. We consider sources emitting intensities which differ for each signal, but remain constant over all rounds of the QKD protocol.

A motivation for this could be the following situation. In the rounds one typically wants to use for the key generation, the so-called signal intensity is sent. Therefore, assume Alice attempts to sent all signals with one equal intensity \(\mu_s\). However, since the setup uses e.g. different laser diodes for each state this assumption often fails and she produces different intensities for each state. Hence, the signal intensity becomes a collection of intensities \(\{\mu_s^x \}_x\) possibly different for each state.

Now, we will show how to calculate asymptotic key rates in this situation. First, start with the signal states in the source replacement scheme including a shield system, generalizing the version used in \cref{sec:Asymptotic Key Rate with Decoy States}.

In the key generation rounds, we can write the state between Alice (\(A\)), her shield system (\(A_s\)) and the signal (\(A'\)) as
\begin{equation}
	\ket{\xi}_{AA_sA'} = \sum_{x,n} \sqrt{\Pr(x,n)} \ket{x}_A \ket{n}_{A_s} \ket{\psi^{(x,n)}}_{A'},
\end{equation}
where \(\Pr(x,n)\) labels the probability of Alice choosing setting \(x\) and sending \(n\)-photons. The probability of sending \(n\) photons implicitly depends on the intensity, thus there is no explicit dependence on the intensity in the equation above. If the signal intensity is equal for all signals \(x\), i.e. \(\mu(x) = \mu_s\) for all signals \(x\), then \(\Pr(x,n)\) simplifies to the usual \(\Pr(x,n) = p_x p_{\mu_s}(n)\). 

Next, we will explore the structure of the state shared by Alice and Bob and relate it to the standard case of equal intensities. For this purpose, Eve can still apply a QND measurement to determine the photon number of Alice's signal, which following \cite{Li2020Phys.Rev.Research} can be recast as an isometry \(V: A' \rightarrow BE\tilde{E} \), where \(E\) and \(\tilde{E}\) are Eve's register and Eve's register storing the photon number, respectively. Furthermore, Eve could always apply a different isometry depending on the photon number she observed, thus \(V\) can be written as a sum over the photon number,
\begin{equation}
	V = \sum_m V_m \Pi_m\otimes \ket{m}_{\tilde{E}},
\end{equation}
where \(V_m\) represents the isometry conditioned on \(m\) photons and \(\Pi_m\) the projector onto \(m\) photons corresponding to Eve's QND measurement. Next, the state after Eve's action is
\begin{equation}
	\begin{aligned}
		\ket{\chi}_{AA_sBE \tilde{E}} = \sum_{x,n,m} &\sqrt{\Pr(x,n)} \ket{x}_A \ket{n}_{A_s} \\
		&\otimes V_m \Pi_m \ket{\psi^{(x,n)}}_{A'} \ket{m}_{\tilde{E}}.
	\end{aligned}
\end{equation}
The projector \(\Pi_m\) acts on \(\ket{\psi^{(x,n)}}_{A'}\) as \(\Pi_m \ket{\psi^{(x,n)}}_{A'} = \delta_{nm} \ket{\psi^{(x,n)}}_{A'}\), and therefore we can simplify \(\ket{\chi}\) to
\begin{equation}
	\begin{aligned}
		\ket{\chi}_{AA_sBE \tilde{E}} = \sum_{x,n} &\sqrt{\Pr(x,n)} \ket{x}_A \ket{n}_{A_s} \\ &\otimes V_m \ket{\psi^{(x,n)}}_{A'} \ket{n}_{\tilde{E}}.
	\end{aligned}
\end{equation}
After tracing out Eve's systems, the state between Alice, Alice's shield and Bob is 
\begin{equation}
	\begin{aligned}
		\rho_{A_sAB} = &\sum_{n=0}^{\infty} \Pr(n) \ketbra{n}_{A_s} \\
		&\otimes \sum_{x,x'} \sqrt{\Pr(x|n) \Pr(x'|n)} \dyad{x}{x'}_A \\
		&\otimes \mathcal{E}_n\left(\dyad{\psi^{(x,n)}}{\psi^{(x',n)}} \right),
	\end{aligned}
\end{equation}
where \(\mathcal{E}_n(\rho) = \Tr_E\left[V_n \rho \left(V_n\right)^{\dagger} \right]\) is Eve's action transforming system \(A'\) to \(B\) as a channel conditioned on \(n\) photons sent by Alice.

Hence, the final state between Alice (both \(A\) and \(A_s\)) and Bob (\(B\)) can still be written as
\begin{equation}\label{eq:State Alice Bob}
	\rho_{A_sAB} = \sum_{n} \Pr(n) \dyad{n}_{A_s} \otimes \rho_{AB}^{(n)},
\end{equation}
which coincides with the results from \cite{Li2020Phys.Rev.Research} if we have equal intensities. However, \(\Pr(n)\) is now the marginal of the joint probability distribution \(\Pr(x,n)\), and \(\Pr(x|n)\) is found by Bayes' theorem,
\begin{equation}
	\Pr(x|n) = \frac{\Pr(n|x) \Pr(x)}{\Pr(n)}.
\end{equation}
Hence, following \cite{Li2020Phys.Rev.Research} the key rate is still
\begin{equation}\label{eq:Key Rate decoy dif int}
	R \geq \sum_{n} \Pr(n) \min_{\rho_{AB}^{(n)} \in S_n} f\left(\rho_{AB}^{(n)}\right) - p_{\text{pass}} \delta_{\text{leak}},
\end{equation}
where \(f(\rho) := D\left(\mathcal{G}\left(\rho \right) || \mathcal{Z}\circ \mathcal{G}\left(\rho \right) \right)\) and \(D(\cdot||\cdot)\) is the relative entropy as in \cite{Winick2018Quantum}.

As in \cref{sec:Asymptotic Key Rate with Decoy States}, the feasible sets \(S_n\) are defined as
\begin{equation}
	\begin{aligned}
		S_n &= \{ \rho_{AB}^{(n)} \in \mathcal{D}(\Hil_A \otimes \Hil_B) \;| \\ 
		&\Pr(x|n) Y^{x,y}_{n,L} \leq \Tr\left[ \Gamma_{x,y} \rho_{AB}^{(n)} \right] \leq \Pr(x|n) Y^{x,y}_{n,U},\\ 
		&\Tr_B(\rho_{AB}) = \sigma_A \}.
	\end{aligned}
\end{equation}

Thus, as before, we are left with the problem of finding the upper and lower bounds on the single photon yields. These can be found by exactly the same SDP as in \cref{eq:SDP decoy} because they are formulated as conditional probabilities (conditioned on Alice sending a particular state).

\section{Flag-state squasher for Passive linear optical Detection setups}\label{sec:Flag-state squasher}
To bring our analysis closer to the experiment, we focus on more general methods for squashing and how to incorporate passive linear optical detection setups into it. This is motivated by the fact that detection setups in reality never posses e.g. a perfect \(50:50\) beamsplitter. However, the simple squashing model from Ref. \cite{Gittsovich2014Phys.Rev.A,Beaudry2008Phys.Rev.Lett.} does not allow for any bias in the detection setup. Therefore, we will use the flag-state squasher of \cite{Zhang2021Phys.Rev.Res.} and extend it to arbitrary passive linear optical detection setups. Again, the aim of this is to allow for a general description of (potentially erroneous) detection setups obeying passive linear optics.

\subsection{Preliminaries}\label{subsec:Preliminaries}
In contrast to our discussion on the unbiased active BB84 protocol, for biased passive six-state and BB84 protocols the squashing map from \cite{Gittsovich2014Phys.Rev.A,Beaudry2008Phys.Rev.Lett.} does not exist. Hence, we will use the alternative of the flag-state squasher, as presented in \cite{Zhang2021Phys.Rev.Res.}. Furthermore, the flag-state squasher is much more flexible and allows the treatment of imperfections in the detection setup unlike the squashing map from \cite{Gittsovich2014Phys.Rev.A,Beaudry2008Phys.Rev.Lett.}, which in general only exists in idealized scenarios. For example, in \cite{Zhang2021Phys.Rev.Res.} the flag-state squasher was applied to a detection efficiency mismatch between detectors.

As mentioned before, the flag-state squasher also applies the mapping
\begin{equation}
	\Lambda: \mathcal{H}_{B_M} \rightarrow \mathcal{H}_{B},
\end{equation}
from the full space \(\mathcal{H}_{B_M}\) to some \emph{finite-dimensional} target space \(\mathcal{H}_{B}\), which is however not the qubit space anymore. Similarly, it is still required that
\begin{equation}
	\Tr[\rho_{\text{in}} F_{B_M}^{(k)} ] = \Tr[\Lambda(\rho_{\text{in}}) F_B^{(k)} ],
\end{equation}
for all input states \(\rho_{\text{in}}\) and POVM elements \(k\). Exploiting the block-diagonal structure of Bob's full POVM, the flag-state squasher leaves the full POVM in the subspace with less than or equal to \(N_B\) photons unchanged and creates the target POVM by replacing all parts with higher photon numbers than \(N_B\) with a flag, i.e.
\begin{equation}
	\begin{aligned}
		F_{B_M}^{(k)} &= F_{B_M,\leq N_B}^{(k)} \oplus F_{B_M,> N_B}^{(k)} \\
		&\mapsto F_B^{(k)} =  F_{B_M,\leq N_B}^{(k)} \oplus \ketbra{k}.
	\end{aligned}
\end{equation}
The flags \(\ketbra{k}\) for each observation \(k\) are orthogonal such that all events corresponding to a photon number above \(N_B\) are known to Eve.

This construction is only useful for QKD if we have a non-trivial bound on the weight inside the \(\leq N_B\) subspace, since Eve could always steer all photons into the \(> N_B\)-photon subspace and Bob would still observe the same statistics. In this case everything about Bob's observations would be known to Eve and one would achieve zero key rate. Therefore, in the remainder of this section we will derive a lower bound on the weight inside the \(\leq N_B\)-photon subspace of Bob's system for passive linear optical detection setups.

\subsection{General Structure of Subspace Estimation}
We will derive an analytical lower bound on the weight inside the \(\leq N_B\)-photon subspace of Bob's system. In contrast to our approach, in \cite{Zhang2021Phys.Rev.Res.} only numerical results were presented. For the lower bound on the weight inside the subspace we will need a suitable observable \(M\) acting on Bob's system, but first, we will discuss in general how one can use any observable to estimate the weight inside the subspace.

For any observable \(M\) on Bob's side, denote by \(M_{\tilde{n}}\) the \(\tilde{n}\)-photon part of \(M\) in terms of the photon number \emph{arriving} at Bob. Due to Eve's interaction this can be different from Alice's photon number \(n\). We find for the measurement outcome \(m_{obs|x}\), conditioned on Alice sending state \(x\) and in terms of the  photon number \(\tilde{n}\) arriving at Bob
\begin{align}
	m_{\text{obs}|x} &=\sum_{\tilde{n}=0}^\infty p(\tilde{n}|x) \Pr \left[m_{\mathrm{obs}} | \tilde{n}, x \right]\\ &= \sum_{\tilde{n}=0}^\infty p(\tilde{n}|x) \Tr \left[ M_{\tilde{n}} \rho^{(\tilde{n})}_x\right].
\end{align}
Next, define \(c_{\tilde{n}} \) as the lower bound on \(\Tr \left[ M_{\tilde{n}} \rho^{(\tilde{n})}\right]\) for all \(\rho^{(\tilde{n})}\) and each \(\tilde{n} \in \N_0\). That means, \(0 \leq c_{\tilde{n}} \leq \min_{\rho^{(\tilde{n})}} \Tr \left[ M_{\tilde{n}} \rho^{(\tilde{n})}\right]\). Thus, the observations \(m_{\text{obs}|x}\) are lower bounded by
\begin{equation}
	m_{\text{obs}|x} \geq \sum_{\tilde{n}=0}^\infty p(\tilde{n}|x) c_{\tilde{n}}.
\end{equation}
Next, one can split this sum into two parts, one involving contributions smaller or equal to Bob's photon number cut-off \(N_B\), i.e. \(\tilde{n} \leq N_B\), and another part for \(\tilde{n} > N_B\). Then, following the approach of \cite[App. A]{Li2020Phys.Rev.Research}, rearranging for \(p(\tilde{n} \leq N_B|x)\) yields a lower bound on the weight inside the subspace
\begin{equation}\label{eq:Lower bound arbitary M}
	\begin{aligned}
		&p(\tilde{n} \leq N_B|x) \geq 1 - \frac{m_{\mathrm{obs}|x} - c_{\leq N_B}}{c_{\geq N_B +1} - c_{\leq N_B}}\\ &=: p_L(\tilde{n} \leq N_B|x),
	\end{aligned}
\end{equation}
where we defined the constants \(c_{\leq N_B}\) and \(c_{\geq N_B +1}\) as
\begin{equation}
		c_{\leq N_B} := \min_{{\tilde{n}}\leq N_B} c_{\tilde{n}}, \qquad 
		c_{\geq N_B +1} := \min_{{\tilde{n}}\geq N_B +1} c_{\tilde{n}}.
\end{equation}
Therefore, given observations \(m_{obs|x}\) and lower bounds \(c_{\tilde{n}}\) we can find a lower bound on the weight inside the \(\leq N_B\)-photon subspace. The task of finding a suitable POVM element \(M\) with constants \(c_{\tilde{n}}\) will be the main challenge of the remainder of this section.

\subsection{Key Rate Optimization Problem including Flag-state Squashing}\label{subsec:Key Rate Optimization Flag-State}
Before, we aim to define a possible POVM element \(M\) for the subspace estimation, let us discuss how one would use a known bound \( p_L(\tilde{n} \leq N_B|n=1)\) on the weight inside the subspace given that Alice sent single photons (\(n=1\)). The exact derivation of \(p(\tilde{n} \leq N_B|x,n)\) from \cref{eq:Lower bound arbitary M} will be stated later.

For now, combining the above bound \cref{eq:Lower bound arbitary M} from the flag-state squasher with the decoy-state key rate in \cref{eq:Key Rate decoy dif int}, we find the following optimization problem for the key rate using only the single photon contribution,
\begin{equation}\label{eq:Key Rate decoy flag}
	\begin{aligned}
	\min_{\rho_{AB}^{(1)} } \; &f\left(\rho_{AB}^{(1)}\right), \\
	\textrm{s.t.}\; &\Pr(x|1) Y^{x,y}_{L,1} \leq \Tr\left[ \Gamma_{x,y} \rho_{AB}^{(1)} \right] \\
	&\hspace{55pt} \leq \Pr(x|1) Y^{x,y}_{U,1}, \; \forall x,y,\\
	&p_L(\tilde{n} \leq N_B|n=1) \leq \Tr[\Pi_{\leq N_B} \rho_{AB}^{(1)}], \\
	&\Tr_B[\rho^{(1)}_{AB}] = \sigma_A  \\
	\end{aligned}
\end{equation}
where \(\sigma_A\) characterizes the constraints originating from the source replacement scheme.

\subsection{Defining a suitable POVM element for Passive Linear Optical Setups}\label{subsec:Defining a suitable POVM element for Passive Linear Optical Setups}
Now, we aim to choose and define a suitable detectable event and associated POVM element \(M\) we use for the subspace estimation only relying on passive linear optics in the detection setup. We assume threshold detectors here, but the analysis can still be used for number resolving detectors, although some changes are required. 

\begin{figure}[ht]
	\centering
	\includegraphics[width=\linewidth]{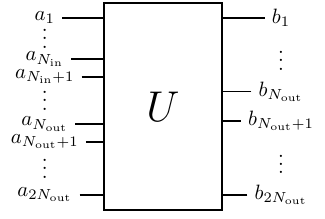}
	\caption{A detection setup using only passive linear optics including \emph{losses} represented as a circuit implementing a linear mode transformation \(U\).}
	\label{fig:Circuit Detector lossy}
\end{figure}

We consider a generic detection setup only comprised of passive linear optical elements. Let us assume, we have \(\Nin\) incoming (non-vacuum) signal modes and \(\Nout\) modes leading to a detection in their respective detector. Furthermore, we model any losses by additional beamsplitters splitting each outgoing mode into two, thus leading to another set of \(\Nout\) modes corresponding to the losses of the respective detector. Hence, our model includes \(\Nin\) incoming signal and \(2 \Nout\) outgoing modes. 

By adding additional \(2\Nout -\Nin\) incoming \emph{vacuum} modes, we can represent the detection setup by a linear mode transformation \cite{Leonhardt2003Rep.Prog.Phys.},
\begin{equation}\label{eq:Linear mode TF}
	b_i = \sum_{j=1}^{2\Nout} U_{ij} a_j,
\end{equation}
where \(\{a_j\}\) are the incoming and \(\{b_i\}\) are the outgoing modes, respectively, and \(U\) is a unitary matrix. A schematic representation of such a mode transformation can be seen in \cref{fig:Circuit Detector lossy}. For example if polarization modes span the input space, then e.g. \(a_H\) and \(a_V\), the annihilation operators of the horizontal and vertical polarization would be the only incoming (non-vacuum) mode operators. 

From here onward, when we refer to the outgoing modes \(b_i\), we implicitly view them as a function of the incoming modes \(a_j\) via the linear mode transformation defined in \cref{eq:Linear mode TF}.

\begin{figure}[ht]
	\centering
	\includegraphics[width=\linewidth]{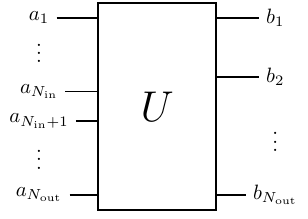}
	\caption{A perfect detection setup using only passive linear optics represented as a circuit implementing a linear mode transformation \(U\).}
	\label{fig:Circuit Detector}
\end{figure}

First, let us assume the case of perfect detectors without any losses. In this case, we can ignore the additional outgoing modes corresponding to losses and also the incoming vacuum modes \(\Nout +1, \dots, 2\Nout\). We label the modes resulting in a detection as \(b_1\) to \(b_{\Nout}\) and the incoming modes by \(a_1\) to \(a_{\Nout}\) out of which \(\Nin\) are non-vacuum modes. This situation is represented in \cref{fig:Circuit Detector}.

Thus, again assuming threshold detectors, a single click in only detector \(i\) in terms of the outgoing modes, is given by the POVM element
\begin{equation}
	F_{\text{sc}}^{i} = \sum_{n=1}^{\infty} \frac{\left(b^{\dagger}_i\right)^n}{n!} \dyad{0} \left(b_i\right)^n,
\end{equation}
and the POVM element of a no-click event in any detector is simply 
\begin{equation}
	F_{\text{no}} = \dyad{0}.
\end{equation}
Note that these POVM elements implicitly depend on the unitary matrix \(U\), as the outgoing modes \(b_i\) are related to the incoming modes via \(U\).

As a first step let us note that only the parts of \(U\) transforming the incoming signal modes matter. For a mathematical justification see \cref{App:Unitary Incoming non-vac modes}.

Therefore, only the first \(\Nin\) columns of \(U\) matter and thus, let us define the \(\Nout \times \Nin\) submatrix \(G\) corresponding to only these columns of \(U\), i.e.
\begin{equation}
	U = \left(
	\begin{matrix}
		\ddots & &  \\
		 & G & \\
		 & & \ddots
	\end{matrix} \, \middle\vert \,
	\begin{matrix}
		\ddots & &  \\
		 & U' & \\
		& & \ddots
	\end{matrix}
	\right).
\end{equation}
Hence, for any \(\rho\)
\begin{equation}\label{eq:trace b using G only}
	\begin{aligned}
		&\Tr \left[ \sum_{i=1}^{\Nout} \frac{\left(b^{\dagger}_i\right)^{\tilde{n}}}{\tilde{n}!} \dyad{0} \left(b_i\right)^{\tilde{n}} \left( \rho \otimes \dyad{0} \right)\right] = \\
		&\sum_{i=1}^{\Nout} \Tr[ \frac{1}{\tilde{n}!} \left(\sum_{j=1}^{\Nin} G_{ij}^* a_j^{\dagger} \right)^{\tilde{n}} \dyad{0} \left(\sum_{k=1}^{\Nin} G_{ik} a_k \right)^{\tilde{n}} \rho ].
	\end{aligned}
\end{equation}

In this perfect case without any losses, we define the POVM element \(M\), used for the lower bound in \cref{eq:Lower bound arbitary M}, as
\begin{equation}\label{eq:POVM flag-state perfect}
	M^{\mathrm{mult}}(G) := \1 - F_{\text{no}} - \sum_{i=1}^{\Nout} F_{\text{sc}}^{i}.
\end{equation}
Due to \cref{eq:trace b using G only} this POVM element effectively only depends on \(G\). For a photon cut-off on Bob's side of \(N_B=1\), i.e. a single photon, the choice of \(M\) is motivated by only being triggered by multi-photon signals. For the subspace estimation, we additionally need the \(\tilde{n}\)-photon part of \(M\) which is given by
\begin{equation}\label{eq:Mn perfect}
	\begin{aligned}
		M_{\tilde{n}}^{\mathrm{mult}}(G) &= \Pi_{\tilde{n}} - \sum_{i=1}^{\Nout} F_{\text{sc}, \tilde{n}}^{i} \\
		&= \Pi_{\tilde{n}} - \sum_{i=1}^{\Nout} \frac{\left(b^{\dagger}_i\right)^{\tilde{n}}}{\tilde{n}!} \dyad{0} \left(b_i\right)^{\tilde{n}}
	\end{aligned}
\end{equation}

Now, the goal is to find a lower bound \(c_{\tilde{n}} \leq \Tr \left[ M_{\tilde{n}} \rho^{(\tilde{n})}\right] \; \rho^{(\tilde{n})} \) by an upper bound on the sum in the previous equation using only the relations between input and output modes of \cref{eq:Linear mode TF}. 

\subsection{Motivating Example: Perfect Passive 4-State Analyzer}\label{subsec:Motivating Example}
As a first example, and to build some intuition what we could hope to achieve, let us consider a perfect passive 4-state receiver. Here, the submatrix \(G\) is given by
\begin{equation}
	G = \begin{pmatrix}
		\sqrt{p_z} & 0 \\
		0 & \sqrt{p_z} \\
		\sqrt{\frac{p_{\text{x}}}{2}} & \sqrt{\frac{p_{x}}{2}} \\
		\sqrt{\frac{p_{\text{x}}}{2}} & -\sqrt{\frac{p_{x}}{2}}
	\end{pmatrix},
\end{equation}
where we ordered the detectors \(1\) to \(4\) such that they correspond to the polarizations \(H,V,D\) and \(A\). Also, note again that due to \cref{eq:trace b using G only} above, each \(\Tr[F_{\text{sc}, n}^{i} \left(\rho \otimes \dyad{0} \right)]\) only depends on row \(i\) of \(G\). Therefore, by knowing \(G\) we can write all single-click probabilities in terms of \(G\). For example we find for detector \(1\) (corresponding to \(H\)),
\begin{equation}
	\begin{aligned}
	&\Tr[F_{\text{sc}, n}^{1} \left(\rho \otimes \dyad{0} \right)] \\
	= &\Tr[ \frac{1}{n!} \left(\sum_{j=1}^{2} G_{1j}^* a_j^{\dagger}\right)^n \dyad{0} \left(\sum_{k=1}^{2} G_{1k} a_k\right)^n \rho ] \\
	=  &\Tr[ \frac{p_z^n}{n!} \left(a_H^{\dagger}\right)^n \dyad{0} \left(a_H\right)^n \rho ] \leq p_z^n.
	\end{aligned}
\end{equation}

The step of minimizing \( \Tr \left[ M_{\tilde{n}} \rho^{(\tilde{n})}\right] \) is equivalent to maximizing the full sum \(\sum_{i=1}^{4} \Tr \left[ F_{\text{sc}, \tilde{n}}^{i}  \rho^{(\tilde{n})}\right]\) over all possible input states \(\rho^{(\tilde{n})}\), however finding the state maximizing the sum is not straightforward. Therefore, as a first step let us maximize each term separately. In this case we find,
\begin{align}
	\max_{\rho^{(\tilde{n})}} \sum_{i=1}^{4} \Tr \left[ F_{\text{sc}, \tilde{n}}^{i}  \rho^{(n)}\right] &\leq \sum_{i=1}^{4} \max_{\rho^{(\tilde{n})}} \Tr \left[ F_{\text{sc}, \tilde{n}}^{i}  \rho^{(n)}\right] \\ 
	&= 2 p_z^{\tilde{n}} + 2 p_x^{\tilde{n}},
\end{align}
because each \(F_{\text{sc}, n}^{i}\) corresponds to a rank one projector. Therefore, the lower bound \(c_{\tilde{n}}\) is
\begin{equation}
	c_{\tilde{n}} = 1 - 2 p_z^{\tilde{n}} - 2 p_x^{\tilde{n}}.
\end{equation}
For \(\tilde{n}=1\) this amounts to \(c_{\tilde{n}} = -1\), which clearly cannot be used for any subspace estimation. Furthermore, for \(p_z=p_x = \frac{1}{2}\) and \(\tilde{n}=2\), we find \(c_{\tilde{n}} = 0\), which gives only trivial bounds in the subspace population estimation. Thus, as one would expect maximizing each term individually will not achieve a useful bound on \(c_{\tilde{n}}\) with small photon numbers.

However, if we partition the sum such that we combine both single-click POVMs in the \(Z\)-basis (\(1\) \& \(2\)) and both single-click POVM elements in the \(X\)-basis (\(3\) \& \(4\)), we have rank two operators and find
\begin{equation}\label{eq:Ex 4-state HV DA}
	\begin{aligned}
		&\max_{\rho^{(n)}} \Tr \left[ \left( F_{\text{sc}, \tilde{n}}^{1} + F_{\text{sc}, \tilde{n}}^{2} \right) \rho^{(\tilde{n})}\right]  \\
		+&\max_{\rho^{(n)}} \Tr \left[ \left( F_{\text{sc}, \tilde{n}}^{3} + F_{\text{sc}, \tilde{n}}^{4} \right) \rho^{(\tilde{n})}\right] = p_z^{\tilde{n}} + p_x^{\tilde{n}}.
	\end{aligned}
\end{equation}
Note that for \(p_z=p_x = \frac{1}{2}\) this amounts to \(c_{\tilde{n}} = 1 -\frac{1}{2^{\tilde{n}-1}} \geq 0\) for all \(\tilde{n} \geq 1\). Inserting this result into \cref{eq:Lower bound arbitary M} and noticing that \(\min_{{\tilde{n}}\leq N_B} c_{\tilde{n}} = c_0 = 0\) yields
\begin{equation}
	p_L(\tilde{n} \leq N_B|x) = 1 - \frac{m_{\mathrm{mult}|x}}{1 - p_z^{N_B +1} - p_x^{N_B +1}},
\end{equation}
which is a useful lower bound on the \(\leq N_B\)-photon subspace for any \(N_B \geq 1\).

Alternatively, one could also partition the sum such that we combine POVM elements from different basis, i.e. (\(1\) \& \(3\)) and (\(2\) \& \(4\)). In this case for simplicity consider \(p_z=p_x = \frac{1}{2}\) only and we find
\begin{equation}
	\begin{aligned}
		&\max_{\rho^{(n)}} \Tr \left[ \left( F_{\text{sc}, \tilde{n}}^{1} + F_{\text{sc}, \tilde{n}}^{3} \right) \rho^{(\tilde{n})}\right]  \\
		+&\max_{\rho^{(n)}} \Tr \left[ \left( F_{\text{sc}, \tilde{n}}^{2} + F_{\text{sc}, \tilde{n}}^{4} \right) \rho^{(\tilde{n})}\right] \\
		&=  \frac{\sqrt{2+\sqrt{2}}^{\tilde{n}}}{2^{\tilde{n}-1}} > \frac{1}{2^{\tilde{n}-1}} .
	\end{aligned}
\end{equation}
This shows, that the previous partition of the sum was indeed better, although chosen completely arbitrary. One fact which distinguished the previous one from this partition mixing the bases was that now the POVM elements in each partition were not orthogonal anymore. Having only orthogonal POVM elements in each sum allows us to detect any multi-photon state having weight in both H and V (or equivalently D and A) at the same time. However, having e.g. the combination \(F_{\text{sc}, \tilde{n}}^{H} + F_{\text{sc}, \tilde{n}}^{D}\) allows for states like \(\ket{1,1}_{HV}\) to still result in a single click with non-vanishing probability and remain unrecognized as a multi-photon state.

For any partitioning of the sum including a higher number of single-click POVM elements than the number of incoming modes (here \(\Nin=2\)), it will in general not be obvious how to estimate a bound analytically. The simple example presented above shows, that although \(M\) is only triggered by multi-photon states the resulting bound on \(c_{\tilde{n}}\) highly depends on the partitioning of the sum, if the full sum cannot be maximized in a single step. Moreover, as indicated by this example, a partitioning by blocks of the number of incoming modes already suffices to achieve a useful lower bound for the subspace estimation.

Now, in order to reach a in general computable upper bound on the sum in \cref{eq:Mn perfect}, we will apply a partitioning with respect to the number of incoming signal modes \(\Nin\), such that each partition of the sum contains at most \(\Nin \) modes. There will be \( k_{\text{max}} = \lfloor \frac{\Nout}{\Nin}\rfloor\) possible blocks involving \(\Nin\) incoming and outgoing modes. If \(\Nout\) is not a multiple of \(\Nin\), then there is also one additional block involving the remaining \(\Nout - k_{\text{max}} \Nin\) outgoing modes.
Let us formalize this as follows, let \(\mathcal{I} \) label the index set \(1, \dots, \Nout\) and let each \(\mathcal{I}_k\), \(k = 1 ,\dots, k_{\text{max}} \), be a subset with \(|\mathcal{I}_k| = \Nin \). Finally, let \(\mathcal{I}_{k_{\text{max}}+1}\) contain \(\Nout - k_{\text{max}} \Nin\) indices. Then,
\begin{equation}
	\mathcal{I} = \bigcup_{k=1}^{k_{\text{max}} +1 } \mathcal{I}_k.
\end{equation}

Connecting this to our previous example of a 4-state receiver, the considered partitionings were
\begin{align}
	\mathcal{I} = \{1,2\} \cup \{3,4\} \quad \text{and} \quad \mathcal{I}' = \{1,3\} \cup \{2,4\},
\end{align}
and it is immediately obvious that there is one more one could consider.

Additionally, each set \(\mathcal{I}_k\) directly corresponds to the rows of \(G\) which are involved in the mode transformation due to \cref{eq:trace b using G only} for that part of the sum. 

This fact allows us to view the results we have shown for the 4-state receiver above from a different perspective as well. For example, each summand in \cref{eq:Ex 4-state HV DA} was in fact the maximum singular value of the corresponding block of \(G\). The same holds true if we chose to partition according to \(\mathcal{I}'\), mixing the bases.

\subsection{Subspace Estimation for Arbitrary Passive Linear Optical Setups}
This intuition can be generalized to an arbitrary \(G\) and the resulting lower bound on the subspace is summarised in the next \cref{Thrm:Lower bound flag state perfect}.

\begin{thrm}[Subspace estimation for perfect passive linear optics detection setup]\label{Thrm:Lower bound flag state perfect}
	Let  \(U \in \C^{\Nout \times \Nout}\) be a unitary matrix describing the mode transformation of a perfect (no losses) passive linear optical setup as in \cref{eq:Linear mode TF}, and label the first \(\Nin\) columns of \(U\) as \(G \in \C^{\Nout \times \Nin}\). By using the POVM element 
	\begin{equation}
		M^{\mathrm{mult}}(G) := \1 - \dyad{0} - \sum_{i=1}^{\Nout} F_{\text{sc}}^{i}.
	\end{equation}
	of \cref{eq:POVM flag-state perfect} and assuming threshold detectors, the \( \tilde{n}\leq N_B \)-subspace can be bounded by
	\begin{equation}
		p_L(\tilde{n} \leq N_B|x) = 1 - \frac{m_{\mathrm{mult}|x}}{c_{\geq N_B +1}},
	\end{equation}
	where \(c_{\geq N_B +1} := \min_{{\tilde{n}}\geq N_B +1} c_{\tilde{n}} \). The constants \(c_{\tilde{n}}\) depend on the partitions \(\mathcal{I} = \bigcup_{k=1}^{k_{\text{max}} +1 } \mathcal{I}_k\), where \( k_{\text{max}} = \lfloor \frac{\Nout}{\Nin}\rfloor\). These partitions are of size \(|\mathcal{I}_k| =\Nin\) for \(k = 1,\dots, \kmax\) and \(|\mathcal{I}_{\kmax +1 }| = \Nout -\kmax \Nin\). The constants \(c_{\tilde{n}}\) are defined as
	\begin{equation}
		\begin{aligned}
			c_{\tilde{n}} &= \max_{\mathcal{I} = \cup \mathcal{I}_k} \left(1 - \sum_{k=1}^{k_{\text{max}}+1} \lambda^{2\tilde{n}}\left(\mathcal{I}_k\right) \right), \quad \tilde{n} > 0, \\
			c_0 &= 0,
		\end{aligned}
	\end{equation}
	where \(\lambda\left(\mathcal{I}_k\right)\) is the maximum singular value of the \(\Nin \times \Nin \) block of \(G\) containing all rows labelled by \(\mathcal{I}_k\).
\end{thrm}
\begin{proof}
	See \cref{App:Proof Subspace Estimation}.
\end{proof}

After covering the perfect detection setup, we consider the situation with losses in the setup. We will reduce this situation to the lossless case by finding an equivalent transformation which first introduces the losses and then acts with an equivalent lossless detection setup onto the states.

As discussed earlier, we model the setup by having \(\Nin\) incoming signal modes and additional \(2\Nout - \Nin\) incoming vacuum modes. The output of the detection setup is then described by \(\Nout\) detection and \(\Nout\) loss modes, as shown in \cref{fig:Circuit Detector lossy}.

Again, note that only the columns in \(U\) corresponding to the transformation of the incoming modes, i.e. \(G\), matter due to \cref{eq:trace b using G only}. Hence, for now view the unitary transformation \(U\) on the modes as a semi-unitary \(G\) (\(G^{\dagger} G = \1_{\Nin} \)) since the vacuum modes do not contribute.

Next, we will decompose the semi-unitary \(G\) into two consecutive ones \(\left( V_d \oplus V_l \right) W\). The first one, \(W\) corresponds to the losses and the remaining ones \(V_d\) and \(V_l\) act independently performing the mode transformation on the detected and loss modes. Then, one can expand those semi-unitaries to unitaries again. The important step is that the extended version of the loss semi-unitary \(W\) will act only as the identity on the incoming vacuum modes \(\Nin +1 , \dots \Nout\). Thus, it does not transform any part of the signal modes into those vacuum modes, such that those will stay in the vacuum state. This decomposition is formalized in the following \cref{Lem:Detector decomposition}.

\begin{lem}[Detector decomposition]\label{Lem:Detector decomposition}
	Let  \(U \in \C^{2\Nout \times 2\Nout}\) be a unitary matrix. Furthermore, label the first \(\Nin\) columns of \(U\) as \(G\). Then, there exist \(V_{d}, V_l \in \C^{\Nout \times \Nin} \) and \(W \in \C^{2\Nin \times \Nin}\) semi-unitaries (\(W^{\dagger} W = \1\)), such that \(G\) can be written as
	\begin{equation}
		G = \left( V_{d} \oplus  V_{l} \right) W.
	\end{equation}
	Additionally, there exists a unitary matrix \(\tilde{U}\) with equal first \(\Nin\) columns \(G\), which can be written as
	\begin{equation}
		\tilde{U}  = \begin{pmatrix}
			T_d & 0 \\
			0 & T_l
		\end{pmatrix} \cdot
		\left(
		\begin{matrix}
			W_d \oplus \frac{\1}{\sqrt{2}} \hfill \\
			W_l \oplus \frac{\1}{\sqrt{2}} 
		\end{matrix}
		\, \middle\vert \,
		W'
		\right) := V_{\text{tot}} \cdot W_{\text{tot}},
	\end{equation}
	where \(W'\) is any matrix completing \(W_{\text{tot}}\) to a unitary, and both \(V_{\text{tot}} \in \C^{2\Nout \times 2\Nout} \) and \( W_{\text{tot}} \in  \C^{2\Nout \times 2\Nout}\) are unitary matrices.
\end{lem}
\begin{proof}
	See \cref{App:Proof Subspace Estimation}.
\end{proof}

Next, to see how we could use this lemma to reduce the loss case to the perfect one and formalize the strategy laid out at the beginning of the section consider the following. Let us label the system of the incoming \(\Nin\) signal modes by \(\text{Sig}\), the system of vacuum modes \(\Nin + 1, \dots \Nout\) by \(\text{Env}_1\) and the remaining incoming modes by \(\text{Env}_2\).

Due to the decomposition of \cref{Lem:Detector decomposition} one can recast the mode transformation by \(U\) as an equivalent transformation due to \(\tilde{U}\) acting equally on the incoming modes in system \(\text{Sig}\). This allows for a decomposition into two steps. Let \(\Omega_{W_{\text{tot}}}\) be the unitary transformation on the Hilbert space implementing the linear mode transformation \(W_{\text{tot}} \). Then, the state in the systems \(\text{Sig},\text{Env}_1\) after the first mode transformation \(W_{\text{tot}}\) is 
\begin{equation}
	\begin{aligned}
		\rho_{\text{Sig},\text{Env}_1} = &\Tr_{\text{Env}_2} \left[ \Omega_{W_{\text{tot}}} \left(\rho_{\text{Sig}} \otimes \ketbra{0}_{\text{Env}_1 \text{Env}_2}\right) 	\Omega_{W_{\text{tot}}}^{\dagger} \right] \\
		= &\sigma_{\text{Sig}} \otimes \ketbra{0}_{\text{Env}_1},
	\end{aligned}
\end{equation}
where \(\sigma_{\text{Sig}} = \Psi(\rho_{\text{Sig}})\) for some CPTP map \(\Psi\), and we used the fact that \(W_{\text{tot}}\) does not transform any modes of system \(\text{Sig}\) into \(\text{Env}_1\) or vice versa. We will give the adversary control over this channel \(\Psi\), which will be formalized in \cref{Thrm:Lower bound flag state lossy}.

By construction, any non-vacuum state in system \(\text{Sig}\) will lead to a detection. Therefore, define new equivalent POVM elements on systems \(\text{Sig},\text{Env}_1\) in terms of the modes transformed due to \(V_d\), resulting from \cref{Lem:Detector decomposition}. The POVM element \(M\) used for the subspace estimation will be written as before, but now depends on \(V_d\), i.e.
\begin{equation}\label{eq:POVM flag-state lossy}
	M^{\mathrm{mult}}(V_d) := \1 - \dyad{0} - \sum_{i=1}^{\Nout} F_{\text{sc}}^{i}.
\end{equation}
It is important to note that each \(b_i\) in the definition of the \(F_{\text{sc}}^{i}\) is now given by
\begin{equation}
	b_i = \sum_{j=1}^{\Nout} (T_d)_{ij} a_j,
\end{equation}
where \(T_d\) is the expanded unitary version of \(V_d\). Again, note that only the first \(\Nin\) columns of \(T_d\) matter due to \cref{eq:trace b using G only}, which is \(V_d\), hence the definition of \(M\) in terms of \(V_d\) only.

Ultimately, we have reduced the loss case to the perfect detector case and can apply the previous bound on \(V_d\), which is summarized in the next \cref{Thrm:Lower bound flag state lossy}.

\begin{thrm}[Subspace estimation for linear optics detection setup with loss]\label{Thrm:Lower bound flag state lossy}
	Let  \(U \in \C^{2\Nout \times 2\Nout}\) be a unitary matrix describing the mode transformation of a passive linear optical setup with losses as in \cref{eq:Linear mode TF}. Let \(V_d \in \C^{\Nout \times \Nin}\) be the semi-unitary resulting from applying \cref{Lem:Detector decomposition} to \(U\). By using the POVM element 
	\begin{equation}
		M^{\mathrm{mult}}(V_d) := \1 - \dyad{0} - \sum_{i=1}^{\Nout} F_{\text{sc}}^{i}.
	\end{equation}
	of \cref{eq:POVM flag-state lossy}, defined in terms of \(V_d\), and assuming threshold detectors, the \( \tilde{n}\leq N_B \)-subspace can be bounded by
	\begin{equation}
		p_L(\tilde{n} \leq N_B|x) = 1 - \frac{m_{\mathrm{mult}|x}}{c_{\geq N_B +1}},
	\end{equation}
	where \(c_{\geq N_B +1} := \min_{{\tilde{n}}\geq N_B +1} c_{\tilde{n}} \). The constants \(c_{\tilde{n}}\) depend on the partitions \(\mathcal{I} = \bigcup_{k=1}^{k_{\text{max}} +1 } \mathcal{I}_k\), where \( k_{\text{max}} = \lfloor \frac{\Nout}{\Nin}\rfloor\). These partitions are of size \(|\mathcal{I}_k| =\Nin\) for \(k = 1,\dots, \kmax\) and \(|\mathcal{I}_{\kmax +1 }| = \Nout -\kmax \Nin\). The constants \(c_{\tilde{n}}\) are defined as
	\begin{equation}
		\begin{aligned}
			c_{\tilde{n}} &= \max_{\mathcal{I} = \cup \mathcal{I}_k} \left(1 - \sum_{k=1}^{k_{\text{max}}+1} \lambda^{2\tilde{n}}\left(\mathcal{I}_k\right) \right), \quad \tilde{n} > 0, \\
			c_0 &= 0,
		\end{aligned}
	\end{equation}
	where \(\lambda\left(\mathcal{I}_k\right)\) is the maximum singular value of the \(\Nin \times \Nin \) block of \(V_d\), containing all rows labelled by \(\mathcal{I}_k\).
\end{thrm}
\begin{proof}
	See \cref{App:Proof Subspace Estimation}.
\end{proof}

\subsection{Including Dark-Counts}

To conclude this section, one would also like to include dark counts in the detection setup. Dark counts can be modelled by a classical post processing. Thus, let \(\{F_{\text{no}}, F_{\text{sc}}^i\} =: \{F_i\}_{i \in \{0,1,\dots, N_{\text{out}}\}}\) be the POVM elements excluding dark counts. These could either originate from a perfect detection setup using \(G\) or from an imperfect setup using \(V_d\) due to \cref{Lem:Detector decomposition}. Furthermore, let \(\{\tilde{F}_i\}_{0,1\dots,\Nout}\) be the POVM elements including dark counts. Hence, the following relation holds that
\begin{equation}\label{eq:PP dark counts}
	\tilde{F}_i = \sum_{j=0}^{N_{\text{out}}} \mathcal{P}_{i|j} F_j,
\end{equation}
where \(\mathcal{P}\) is the relevant part of the stochastic matrix implementing the post processing corresponding due to dark counts. Thus, it holds \(\sum_{i} \mathcal{P}_{ij} = p_j \; \forall j\), which we will use in the following corollary. Furthermore, \cref{Cor:Lower bound including dark counts} elevates the previous lower bounds to one which can also be used for detectors with dark counts.

\begin{cor}[Lower bound including dark counts.]\label{Cor:Lower bound including dark counts}
	Let \(G \in \C^{\Nout \times \Nin}\) be the submatrix describing a detection setup consisting only of passive linear optical elements. Furthermore, let \(\mathcal{P}\) be the stochastic matrix implementing the post processing corresponding to dark counts as in \cref{eq:PP dark counts}. Then, the lower bounds on the subspaces of the previous \cref{Thrm:Lower bound flag state perfect} and \cref{Thrm:Lower bound flag state lossy} are still valid using observations including dark counts and the constants \(c_{\tilde{n}}\) from the setups without dark counts.
\end{cor}
\begin{proof}
	See \cref{App:Proof Subspace Estimation}.
\end{proof}

Again, let us point out that the bound resulting from \cref{Cor:Lower bound including dark counts} is strictly worse than \cref{Thrm:Lower bound flag state perfect} or \cref{Thrm:Lower bound flag state lossy}, as the dark counts do not decrease the lower bounds on the \(c_{\tilde{n}}\), but they increase the observation \(m_{\text{obs}}\), thus decreasing the lower bound. 

Finally, the last remaining question is how to find the detector matrix \(G\) for an experimental setup. We will discuss this question in more detail in \cref{App:Detector Characterization for Flag-State Squasher}.

\subsection{Resulting Subspace Bounds for Decoy-State Methods}
To summarize our steps taken so far, by defining the specific POVM element \(M\) we found a generic bound on the weight inside Bob's \(\leq N_B\)-photon subspace. This allows us to include all kinds of detection setups which are governed by passive linear optics. Especially, we can consider different detection efficiencies in each detector and dark counts.

Finally, to formulate our decoy-state analysis and evaluate the key rate optimization \cref{eq:Key Rate decoy flag}, we need a lower bound on the weight in Bob's subspace conditioned on Alice sending \(n\) photons, i.e. \(p(\tilde{n} \leq N_B|x,n)\), depending only on the observations \(m_{\text{mult}|x}\) instead of \(m_{\text{mult}|x,n}\), which cannot be observed. This bound can be reached in two separate ways. First, the simple version can be found by considering
\begin{align}
	&1 - \frac{m_{\mathrm{mult}|x}}{c_{\geq N_B +1}} = p_L(\tilde{n} \leq N_B|x) \\
	&\leq p(\tilde{n} \leq N_B|x) = \sum_{m=0}^{\infty} p(m|x) p(\tilde{n}\leq N_B |x,m) \\ 
	&\leq p(n|x) p(\tilde{n} \leq N_b | x,n) + \left( 1 - p(n|x)\right),
\end{align}
Here, the first equality is by the definition of the lower bounds from \cref{Thrm:Lower bound flag state perfect} or \cref{Thrm:Lower bound flag state lossy}, the second equality is due to chain rules of probabilities, and the last inequality is by bounding the sum with one specific \(n\). Then, solving for \(p(\tilde{n} \leq N_B | x,n)\) yields
\begin{equation}\label{eq:Lower bound NB conditioned on n}
	\begin{aligned}
		&p_L(\tilde{n} \leq N_B|x,n) = 1 - \frac{m_{\text{mult}|x}}{p(n|x) \left( 1 - c_{\geq N_B +1} \right)}.
	\end{aligned}
\end{equation}
This is the lower bound on the subspace population we will use in the formulation of the decoy analysis.

However, one can also use the result of the decoy analysis to construct the lower bound conditioned on \(n\) photons sent by Alice and use that in the minimization of the secret key rate, \cref{eq:Key Rate decoy flag}. If we identify \(m_{\text{obs}|x,n}^U\) with the upper bound on \(m_{\text{obs}|x,n}\) resulting from the decoy analysis, a tighter bound on the subspace population is
\begin{equation}
	\begin{aligned}
		&p_L(\tilde{n} \leq N_B|x,n) = 1 - \frac{m_{\text{mult}|x,n}^U}{ 1 - c_{\geq N_B +1} },
	\end{aligned}
\end{equation}
which we will use for the key rate optimization problem in \cref{eq:Key Rate decoy flag}.

Now, it is important to compare our results with the ones found in Ref.~\cite{Zhang2021Phys.Rev.Res.}. First, the lower bounds in Ref.~\cite{Zhang2021Phys.Rev.Res.} were found by running a numerical experiment for a specific setup with a specifically chosen POVM element. Calculating the bounds numerically guarantees finding the optimal lower bound, but has a limit on Bob's subspace size \(N_B\) (\(N_B = 20\) in their case). Furthermore, the numerical results only show that these lower bounds are monotonically increasing up to \(N_B = 20\), which however is not sufficient for a QKD security proof.

On the other hand, our results here give a generic lower bound which is monotonically increasing in \(N_B\) and do not require designing a specific POVM element for each detection setup. Thus, the results presented here can be used in a QKD security proof and are far more convenient to apply.

\section{Example 2: Biased passive WCP 6-state}\label{sec:Example 2: Biased passive 6-state}
As another example apart from the BB84 protocol, we present the six-state protocol because here 
the squashing map from \cite{Gittsovich2014Phys.Rev.A,Beaudry2008Phys.Rev.Lett.} exists only without any bias in the basis choice. A bias is advantageous for the final secret key rate, but can also originate from beamsplitters having an imperfect splitting ratio. Therefore, we will use the flag-state squasher derived in the previous \cref{sec:Flag-state squasher}. This illustrates the modularity of our approach and allows us to use a bias in Bob's basis choices which is not possible with the simple squashing method from \cite{Gittsovich2014Phys.Rev.A,Beaudry2008Phys.Rev.Lett.}. Furthermore, we will show how to mitigate the problem of different intensities for each signal being sent.

\subsection{Flag-state squasher}
\begin{figure}[hb]
	\centering
	\includegraphics[width=\linewidth]{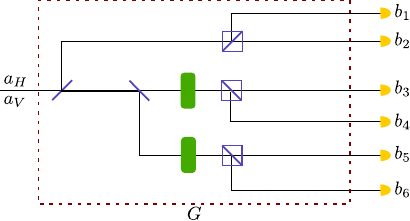}
	\caption{Example of a perfect passive detection setup for a six-state analyser. We label beamsplitters with blue lines, polarizing beamsplitters with additional blue boxes, polarization rotators with green boxes and detectors with yellow paraboloids. The submatrix \(G\) includes all elements in the red dashed box to a semi-unitary.}
	\label{fig:Detector 6 state}
\end{figure}

For this specific example, we assume a perfect passive six-state analyser without any losses and use the POVM elements from \cite{Gittsovich2014Phys.Rev.A,Beaudry2008Phys.Rev.Lett.}. A schematic representation of a six-state receiver including our submatrix \(G\) can be seen in \cref{fig:Detector 6 state}.
If we order \(H,V,D,A,R\) and \(L\) as \(1\) to \(6\) we find the detector matrix \(G\) to be,
\begin{equation}
	G = \begin{pmatrix}
		\sqrt{p_{\text{HV}}} & 0 \\
		0 & \sqrt{p_{\text{HV}}} \\
		\sqrt{\frac{p_{\text{AD}}}{2}} & \sqrt{\frac{p_{\text{AD}}}{2}} \\
		\sqrt{\frac{p_{\text{AD}}}{2}} & -\sqrt{\frac{p_{\text{AD}}}{2}} \\
		\sqrt{\frac{p_{\text{RL}}}{2}} & i\sqrt{\frac{p_{\text{RL}}}{2}} \\
		\sqrt{\frac{p_{\text{RL}}}{2}} & -i\sqrt{\frac{p_{\text{RL}}}{2}} \\
	\end{pmatrix}.
\end{equation}
As we have already laid out in \cref{subsec:Preliminaries}, one needs to find a lower bound for the weight in Bob's \(\leq N_B\)-photon subspace. Since we have a perfect detection setup, we can apply \cref{Thrm:Lower bound flag state perfect} and find
\begin{equation}
	p_L(\tilde{n} \leq N_B|x) = 1 - \frac{m_{\mathrm{mult}|x}}{c_{\leq N_B+1}}, 
\end{equation}
where 
\begin{equation}
	c_{\leq N_B+1} = 1- \sum_{\alpha \in\{\text{HV}, \text{AD}, \text{RL}\}} p_{\alpha}^{N_B+1},
\end{equation}
and \(p_{\alpha}\) is Bob's probability of choosing basis \(\alpha\). As side note it is worth mentioning that in this case as in the motivating example \cref{subsec:Motivating Example}, the optimal partition of \cref{Thrm:Lower bound flag state perfect} of \(G\) is achieved by arranging all rows in \(G\) of the same detection basis together.

Furthermore, applying \cref{eq:Lower bound NB conditioned on n} for the weight inside the subspace, conditioned on state \(x\) sent by Alice with \(n\) photons, we find
\begin{equation}\label{eq:6state Lower bound NB conditioned on n}
	\begin{aligned}
		&p_{L}^{\text{6-state}}(\tilde{n} \leq N_B|x,n) \\
		&= 1 - \frac{m_{\mathrm{mult}|x}}{p(n|x) \left(1-\sum_{\alpha} p_{\alpha}^{N_B+1} \right)}.
	\end{aligned}
\end{equation}
Then the bound on the subspace without conditioning on Alice's state is the marginal, i.e.
\begin{equation}
	\begin{aligned}
		&p_{L}^{\text{6-state}}(\tilde{n} \leq N_B|n) \\
		&= \sum_x p(x|n) 	p_{L}^{\text{6-state}}(\tilde{n} \leq N_B|x,n)
	\end{aligned}
\end{equation}
Now, we have all necessary tools to formulate both our improved decoy methods and the final key rate optimization.

\subsection{Decoy analysis}
With our squashing map and the subspace estimation at hand, one can state the minimization problem for the \(n\)-photon yields for the 6-state receiver as
\begin{equation}\label{eq:SDP 6-state}
	\begin{aligned}
		Y_{n,L}^{i,j}:=&\min_{\mbf{Y}_m, J_0, J_1}\;   Y^{i,j}_{n}\\
		\textrm{s.t.}\; &  \gamma^{\mu}_{y|x} \leq \sum_{m\leq N_{\mathrm{ph}}} p_{\mu}(m)\; Y_m^{x,y} +(1-p_{tot}), \\
		&\gamma^{\mu}_{y|x} \geq \sum_{m\leq N_{\mathrm{ph}}} p_{\mu}(m)\; Y_m^{x,y}, \\
		&Y_m^{x,y} = \Tr\left[J_m \left(F_y^B \otimes \left(\rho_x^{(m)}\right)^T \right)\right], \\
		&0 \leq Y^{x,y}_m \leq 1, \; \forall \; 0 \leq m \leq N_{\mathrm{ph}}, \\
		& 1 \geq \Tr\left[J_m \left(\Pi_{\leq N_B} \otimes \left(\rho_x^{(m)}\right)^T \right)\right] \\
		&\hspace{5pt} \geq p_L(\tilde{n} \leq N_B|x,m), \; m=0,1,\\
		&J_0,\; J_1 \succeq 0, \Tr_B\left[J_0 \right] = 1, \; \Tr_B\left[J_1 \right] = \1_{A'}, \\
		&\forall \mu \in \{\mu_1, \mu_2, \dots\}, \; \forall x,y,
	\end{aligned}    
\end{equation}
where \( \Pi_{\leq N_B}\) is again the projector onto Bob's subspace with \(\tilde{n} \leq N_B\) photons and \(p_L\) is the lower bound from \cref{eq:6state Lower bound NB conditioned on n}. Bob's POVM elements \(F_y^B\) acting on the \(\leq N_B\)-photon subspace are presented in \cref{App:POVMs and Kraus ops 6-state}.

\subsection{Final Key Rate Optimization}
Next, using the bounds on single photon yields resulting from \cref{eq:SDP 6-state} and the bound for the flag-state squasher, we can formulate the final key rate optimization problem. Following \cref{subsec:Key Rate Optimization Flag-State} we find the final optimization to be
\begin{equation}
	\begin{aligned}
		\min_{\rho_{AB}^{(1)} } \; &f\left(\rho_{AB}^{(1)}\right), \\
		\textrm{s.t.}\; &\Pr(x|1) Y^{x,y}_{L,1} \leq \Tr\left[ \Gamma_{x,y} \rho_{AB}^{(1)} \right] \\
		&\hspace{55pt} \leq \Pr(x|1) Y^{x,y}_{U,1}, \; \forall x,y\\
		&\Tr_B[\rho^{(1)}_{AB}] = \sigma_A,  \\
		&\Tr[\Pi_{\leq N_B} \rho_{AB}^{(1)}] \geq p_{L}^{\text{6-state}}(\tilde{n} \leq N_B|n=1).
	\end{aligned}
\end{equation}

\subsection{Results with equal Intensities}\label{sec:6 state results}
In this section, we will compare the results of our improved decoy-state methods with previous results and furthermore showcase the performance of our flag-state squashing methods. For all of our results in this section we again assume a loss parameter of \(\unit[0.2]{dB/km}\).

\begin{figure}[hb]
	\centering
	\includegraphics[width=\linewidth]{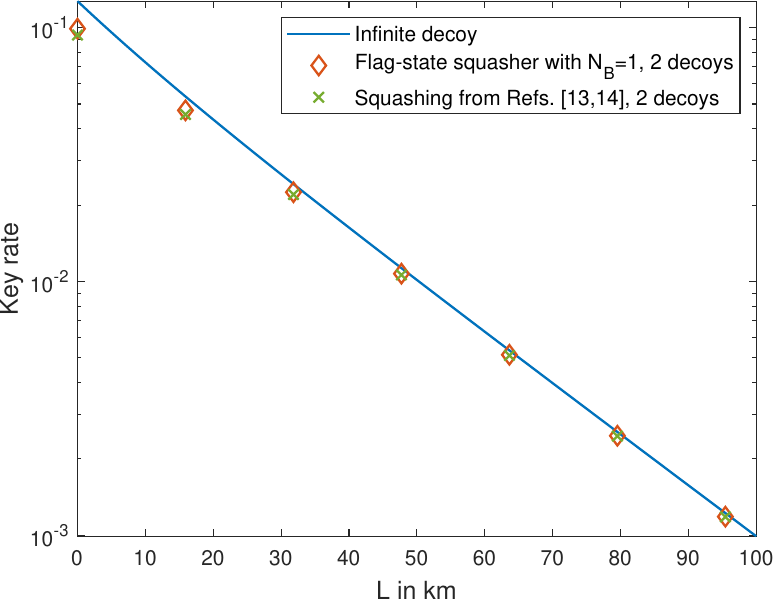}
	\caption{Comparison of secret key rates of the passive unbiased six-state protocol over the distance between numerical results for three intensities using the flag-state squasher with \(N_B=1\) presented in this work (orange diamonds) and results from \cite{Gittsovich2014Phys.Rev.A,Beaudry2008Phys.Rev.Lett.} (green crosses).}
	\label{fig:6-state comparison flagstate unbiased}
\end{figure}

First, in \cref{fig:6-state comparison flagstate unbiased} we present a comparison between the flag-state squasher with \(N_B=1\) and the squashing map from \cite{Gittsovich2014Phys.Rev.A,Beaudry2008Phys.Rev.Lett.}. Here we used equal basis choices, i.e. \(p_x = p_y = p_z = \frac{1}{3}\), since otherwise the squashing map from \cite{Gittsovich2014Phys.Rev.A,Beaudry2008Phys.Rev.Lett.} does not exist. Furthermore, since this comparison is about the squashing methods, we chose \(3\) intensities such that there is almost no influence from the decoy-state methods. Moreover, we used the simple linear program for the yield estimation from \cref{eq:LP decoy} for both curves. For the error correction we assumed the Shanon limit with an efficiency of \(f_{\text{EC}} =1\). As one would hope, the results are mostly equal and thus, the flag-state squasher recovers previous results already for small subspace dimensions. This means in practice one can replace the simple squashing map from \cite{Gittsovich2014Phys.Rev.A,Beaudry2008Phys.Rev.Lett.} with the flag-state squasher, to achieve higher flexibility in terms of the detection setups.

\begin{figure}[ht]
	\centering
	\includegraphics[width=\linewidth]{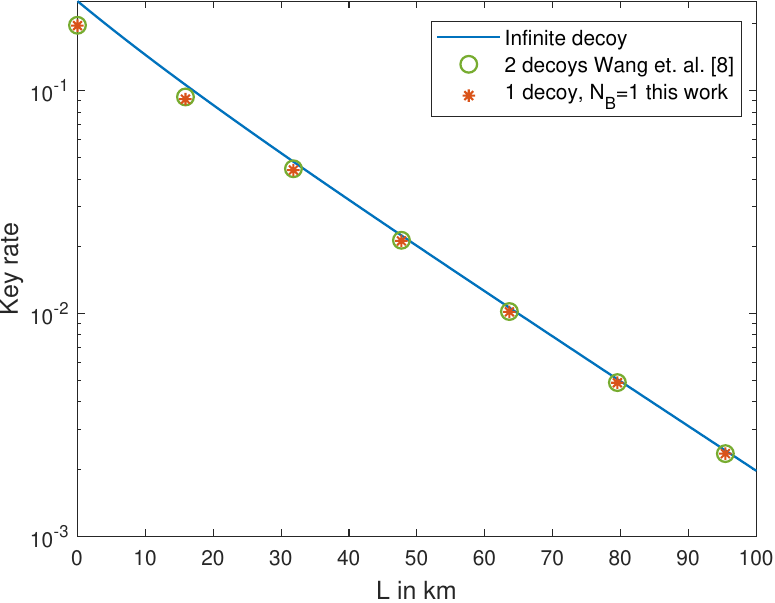}
	\caption{Comparison of secret key rates of the passive biased (\(p_z =0.8\), \(p_x=p_y=0.1\)) six-state protocol over the distance between our numerical results for one decoy intensity (orange stars) with numerical results from \cite{Wang2022Phys.Rev.Res.} using two decoy intensities (green circles).}
	\label{fig:6-state comparison flagstate biased}
\end{figure}

Next, in \cref{fig:6-state comparison flagstate biased} we show that even while using the flag-state squasher the same increases in key rates originating from the improved constraints in the decoy analysis, \cref{eq:SDP decoy}, apply. In particular, we compare the different decoy-state methods, i.e. the linear program compared to our improved decoy methods. Since the flag-state squasher allows us to consider a biased detection setup, we chose \(p_z = 0.8\), \(p_x=p_y = 0.1\). The error correction efficiency is assumed to be the same \(f_{\text{EC}} =1 \) as for the previous example of BB84. Again, our improved method of \cref{eq:SDP decoy} recovers the same key rate as the linear program from \cite{Wang2022Phys.Rev.Res.} with only two intensities just as for the BB84 protocol.

\begin{figure}[ht]
	\centering
	\includegraphics[width=\linewidth]{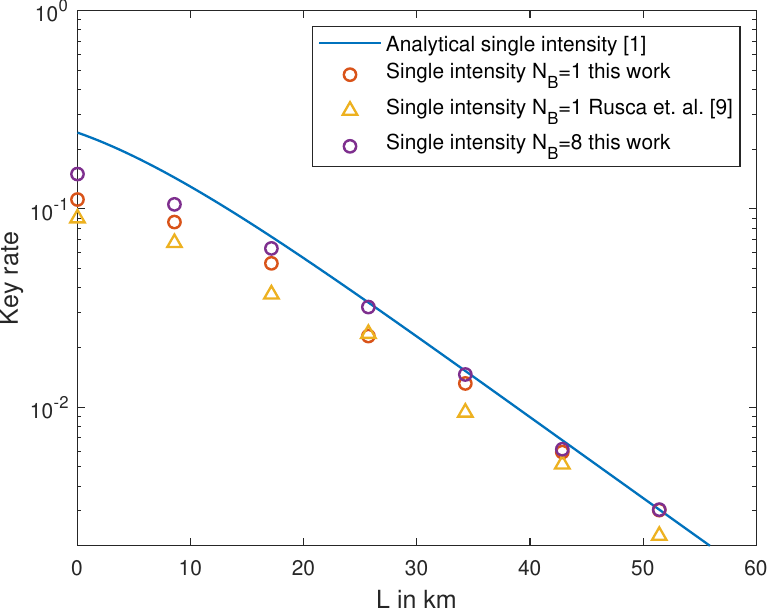}
	\caption{Comparison of secret key rates of the 6-state protocol without decoy intensities over distance between our numerical results with \(N_B=1\) (orange circles), \(N_B=8\) (purple circles), numerical results from \cite{Rusca2018Appl.Phys.Lett.} (yellow triangles) and analytical results from \cite{Lutkenhaus2000Phys.Rev.A} (blue line).}
	\label{fig:6-state flagstate single}
\end{figure}

Finally, in \cref{fig:6-state flagstate single} we also consider the no-decoy case for the six-state protocol. Without any decoy intensities and a flag dimension of \(N_B=1\) we do not quite recover the analytical key rate of Ref. \cite{Lutkenhaus2000Phys.Rev.A}. This can be seen in \cref{fig:6-state flagstate single} by the blue solid curve and the red circles. 

The reason for this is the remaining looseness in the flag-state squasher, not our improved decoy methods. However, we can recover the results from \cite{Lutkenhaus2000Phys.Rev.A} by increasing \(N_B\). In \cref{fig:6-state flagstate single} one can see an example for \(N_B = 8\) (purple circles), which already shows a significant improvement over \(N_B=1\). Thus, indeed, the lower bound from the flag-state squasher is the reason for the discrepancy. We expect that the limit of \(N_B \rightarrow \infty\) will recover the analytical key rate as \(p_L \overset{N_B \rightarrowtail \infty}{\longrightarrow} 1\). This looseness in the flag-state squasher is unimportant as soon as we introduce decoy states, because the bounds on the yields become much tighter.

Furthermore, in \cref{fig:6-state flagstate single} we also show the equivalent results only imposing \(e_0 = \frac{1}{2}\) as in \cite{Rusca2018Appl.Phys.Lett.} (yellow triangles). Again, our methods show significantly better results compared to this simple case, which would not recover the analytical key rates even in the limit \(N_B \rightarrow \infty\).

\subsection{Results with Differing Intensities}
At last, we show how the case of differing intensities changes the asymptotic key rates and apply our results from \cref{sec:Extension of Asymptotic Key Rate with Differing Intensities}. Having different intensities for each state effectively changes the probabilities each signal is sent with conditioned on a single photon. Most importantly, it also introduces an effective bias in the bit choices which can be different for each basis. We will explain two options how to resolve this situation.

Broadly speaking, there are two extreme cases, one can keep the protocol as it is and accept the changed probabilities or actively change the bit bias and counteract the effect.
For the latter case of applying a bit bias, one would aim at
\begin{equation}
	\frac{1}{2} \overset{!}{=} \Pr(x|n) = \frac{\Pr(n|x) \Pr(x)}{\Pr(n)},
\end{equation}
which means that signals with higher intensities are sent less often. This already shows one problem of this approach, as the resulting probability of sending a single photon might be reduced. Thus, the prefactor \(\Pr(1)\) in the key rate formula \cref{eq:Key Rate decoy dif int} will be smaller. While keeping the relative entropy term constant, this will effectively reduce the total key rate.

On the other hand, if we just accept the effective bit bias from the differing intensities, the prefactor \(\Pr(1)\) will stay constant. However, the amount of secret key we can generate per single photon will be smaller, because the effective bit bias will reduce the relative entropy in \cref{eq:Key Rate decoy dif int}, which is the disadvantage of applying no countermeasure.

\begin{figure}[hb]
	\centering
	\includegraphics[width=\linewidth]{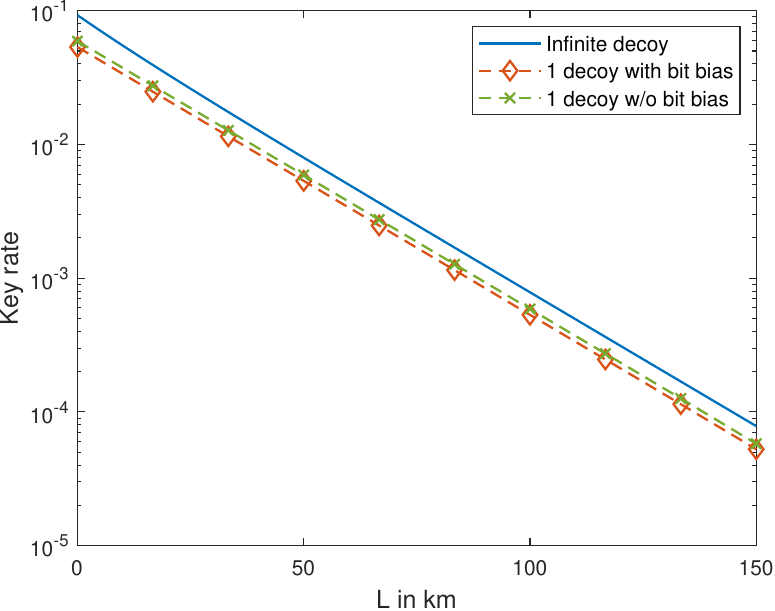}
	\caption{Comparison of secret key rates of the 6-state protocol with one decoy intensity over the distance between the rates using a bit bias (red diamonds) and no bit bias (green crosses).}
	\label{fig:6statedifint}
\end{figure}

Thus, one has two competing effects, but the resulting key rates can be very similar. To illustrate this, we calculated the asymptotic key rates for a signal intensity of \(\mu_{\text{sig}} = (0.95,0.15,0.1,0.9,0.95,0.1)\), which is ordered in terms of the states H, V, etc. The equivalent effective signal intensity (weighted average of the differing intensities) with equal single-photon probability is \(\bar{\mu}_{\text{sig}} = 0.3256\). For simplicity, we kept the second intensity equal for all states, i.e. \(\mu_2=0.01\), since it is only used in the yield estimation. 

\begin{figure}[ht]
	\centering
	\includegraphics[width=\linewidth]{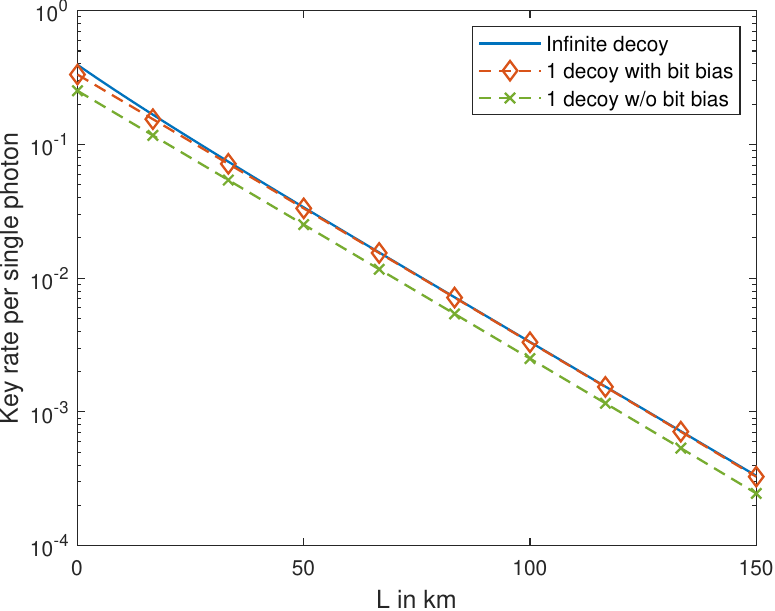}
	\caption{Comparison of secret key rates \emph{per single photon} of the 6-state protocol with one decoy intensity over the distance between the rates using a bit bias (red diamonds) and no bit bias (green crosses).}
	\label{fig:6statedifintpersinglephoton}
\end{figure}

In \cref{fig:6statedifint} one can see the resulting secret key rates with and without a bit bias. Both approaches result in a very similar secret key rate, but neither can reach the key rate of infinite decoy with an equal signal intensity of \(\bar{\mu}_{\text{sig}} = 0.3256\). The curve without a compensating bit bias (green crosses), is having a lower key rate because again, the different intensities for each signal lead to a resulting bias, ultimately lowering the key rate. On the other hand, the curve with a bit bias (red diamonds), cancelling the overall bias, has a lower key rate than an equivalent decoy protocol with a signal intensity of \(\mu_{\text{sig}} = \bar{\mu}_{\text{sig}} = 0.3256\). This happens, since \(p_{\bar{\mu}_{\text{sig}}}(1) > \Pr(n=1)\), as also discussed before.

However, if we consider the key rate per single photon, i.e. 
\begin{equation}
	\frac{R_{\infty}}{\Pr(1)},
\end{equation}
as shown in \cref{fig:6statedifintpersinglephoton}, then we can see that the protocol compensating the bit bias is actually recovering the same key rate per single photon as a protocol with equal intensities.

Ultimately, there will be a trade-off between compensating the bias due to differing intensities and keeping the probability \(\Pr(n=1)\) of Alice sending single photon as high as possible. Therefore, one should optimize Alice's probabilities of sending a state, however, as we have seen the potential increase in key rate is rather small. 

\section{Conclusion}\label{sec:Conclusion}
We have presented improved decoy-state methods which tighten the achievable key rates in two ways. First, they recover the analytical no-decoy rates \cite{Lutkenhaus2000Phys.Rev.A} and second, they improve the key rates from the standard linear program approach. As in \cite{Rusca2018Appl.Phys.Lett.} we see that a single decoy intensity suffices for experimental implementations with our improved decoy methods, while still slightly improving on their results. Moreover, we reach higher key rates than previous results \cite{Wang2022Phys.Rev.Res.} using two decoy intensities, which is essentially a free gain in key rate by a more refined analysis. Finally and most importantly, we reach tight results for the no-decoy case opposed to \cite{Rusca2018Appl.Phys.Lett.}.

In addition, we presented new analytical bounds for the flag-state squasher for arbitrary passive linear optical detection setups. This new analytical bound allow us to include for example bias in the detection basis, dark counts and different detection efficiencies. Most importantly, it removes the previous numerical assumption of the monotonicity in the lower bounds.

Furthermore, we presented a key rate formula applicable to differing intensities for each signal and showed results for the six-sate protocol. We presented two strategies for the situation when the intensities vary widely, one of which is applying an additional bias to the bit values, counteracting the effective bias by the differing intensities.

\section*{Acknowledgements}
We would like to especially thank Shlok Nahar and John Burniston for the enlightening discussions about decoy-state methods and the weight estimation for the flag-state squasher.
 
The research has been conducted at the Institute for Quantum Computing, at the University of Waterloo, which is supported by Innovation, Science, and Economic Development Canada. Support was also provided by NSERC under the Discovery Grants Program, Grant No. 341495.

\bibliography{ref, poster}

\appendix
\section{Unitary matrix of Detection setup only depends on incoming non-vacuum modes}\label{App:Unitary Incoming non-vac modes}
As described in \cref{subsec:Defining a suitable POVM element for Passive Linear Optical Setups} only the parts of \(U\) transforming the incoming signal modes matter. To see this, consider only \(\Tr[\frac{\left(b^{\dagger}_i\right)^{\tilde{n}}}{\tilde{n}!} \dyad{0} \left(b_i\right)^{\tilde{n}} \left(\rho \otimes \dyad{0} \right) ]\) for an arbitrary \(\rho\) defined only on the incoming signal modes. If we expand this expression using \cref{eq:Linear mode TF} and the multinomial theorem we find
\begin{align}
	&\Tr[\frac{\left(b^{\dagger}_i\right)^{\tilde{n}}}{\tilde{n}!} \dyad{0} \left(b_i\right)^{\tilde{n}} \left(\rho \otimes \dyad{0} \right) ]\\
	&= \Tr[ \frac{1}{\tilde{n}!} \left(\sum_{j=1}^{\Nout} U_{ij}^* a_j^{\dagger} \right)^{\tilde{n}} \dyad{0} \left(\sum_{k=1}^{\Nout} U_{ik} a_k \right)^{\tilde{n}} \left(\rho \otimes \dyad{0} \right) ] \\ 
	&= \frac{1}{\tilde{n}!} \sum_{\substack{k_{1}+\dots +k_{\Nout}=\tilde{n} \nonumber \\ l_{1}+\dots +l_{\Nout}=\tilde{n}}} {\tilde{n} \choose k_{1},\ldots ,k_{\Nout}} {\tilde{n} \choose l_{1},\ldots ,l_{\Nout}} \\ 
	\cdot &\Tr[ \prod_{\substack{t=1, \\ s=1}}^{\Nout}  \left(U_{it}^* a_{t}^{\dagger}\right)^{k_t} \dyad{0} \left(U_{is} a_s\right)^{l_s} \left(\rho \otimes \dyad{0} \right) ] \\ 
	&= \sum_{\substack{k_{1}+\dots +k_{\Nout}=\tilde{n} \\ l_{1}+\dots +l_{\Nout}=\tilde{n}}} \sqrt{\tilde{n} \choose k_{1},\ldots ,k_{\Nout}} \sqrt{\tilde{n} \choose l_{1},\ldots ,l_{\Nout}} \nonumber \\ 
	\cdot &\Tr[ \prod_{\substack{t=1, \\ s=1}}^{\Nout}  \left(U_{it}^*\right)^{k_t} U_{is}^{l_s} \dyad{k_1,\dots,k_{\Nout}}{l_1,\dots,l_{\Nout}} \left(\rho \otimes \dyad{0} \right) ] \\
	&= \Tr[ \frac{1}{\tilde{n}!} \left(\sum_{j=1}^{\Nin} U_{ij}^* a_j^{\dagger} \right)^{\tilde{n}} \dyad{0} \left(\sum_{k=1}^{\Nin} U_{ik} a_k \right)^{\tilde{n}} \rho ]
\end{align}
Hence, for any \(\rho\)
\begin{equation}
	\begin{aligned}
		&\Tr \left[ \sum_{i=1}^{\Nout} \frac{\left(b^{\dagger}_i\right)^{\tilde{n}}}{\tilde{n}!} \dyad{0} \left(b_i\right)^{\tilde{n}} \left( \rho \otimes \dyad{0} \right)\right] = \\
		&\sum_{i=1}^{\Nout} \Tr[ \frac{1}{\tilde{n}!} \left(\sum_{j=1}^{\Nin} G_{ij}^* a_j^{\dagger} \right)^{\tilde{n}} \dyad{0} \left(\sum_{k=1}^{\Nin} G_{ik} a_k \right)^{\tilde{n}} \rho ],
	\end{aligned}
\end{equation}
as stated in the main text.

\section{Proofs of Subspace Estimation for perfect and imperfect Detection Setups}\label{App:Proof Subspace Estimation}
\setcounter{defn}{0}

\begin{thrm}[Subspace estimation for perfect passive linear optics detection setup (restated)]
	Let  \(U \in \C^{\Nout \times \Nout}\) be a unitary matrix describing the mode transformation of a perfect (no losses) passive linear optical setup as in \cref{eq:Linear mode TF}, and label the first \(\Nin\) columns of \(U\) as \(G \in \C^{\Nout \times \Nin}\). By using the POVM element 
	\begin{equation}
		M^{\mathrm{mult}}(G) := \1 - \dyad{0} - \sum_{i=1}^{\Nout} F_{\text{sc}}^{i}.
	\end{equation}
	of \cref{eq:POVM flag-state perfect} and assuming threshold detectors, the \( \tilde{n}\leq N_B \)-subspace can be bounded by
	\begin{equation}
		p_L(\tilde{n} \leq N_B|x) = 1 - \frac{m_{\mathrm{mult}|x}}{c_{\geq N_B +1}},
	\end{equation}
	where \(c_{\geq N_B +1} := \min_{{\tilde{n}}\geq N_B +1} c_{\tilde{n}} \). The constants \(c_{\tilde{n}}\) depend on the partitions \(\mathcal{I} = \bigcup_{k=1}^{k_{\text{max}} +1 } \mathcal{I}_k\), where \( k_{\text{max}} = \lfloor \frac{\Nout}{\Nin}\rfloor\). These partitions are of size \(|\mathcal{I}_k| =\Nin\) for \(k = 1,\dots, \kmax\) and \(|\mathcal{I}_{\kmax +1 }| = \Nout -\kmax \Nin\). The constants \(c_{\tilde{n}}\) are defined as
	\begin{equation}
		\begin{aligned}
			c_{\tilde{n}} &= \max_{\mathcal{I} = \cup \mathcal{I}_k} \left(1 - \sum_{k=1}^{k_{\text{max}}+1} \lambda^{2n}\left(\mathcal{I}_k\right) \right), \quad n\geq 0, \\
			c_0 &= 0,
		\end{aligned}
	\end{equation}
	where \(\lambda\left(\mathcal{I}_k\right)\) is the maximum singular value of the \(\Nin \times \Nin \) block of \(G\) containing all rows labelled by \(\mathcal{I}_k\).
\end{thrm}
\begin{proof}
	For simplicity, consider the trivial partition
	\begin{align}
		&\begin{aligned}
		\mathcal{I} = &\{1, \dots, \Nin\} \cup \dots \\
		&\cup \{\left(k_{\text{max}} - 1\right)\Nin +1 ,\dots, k_{\text{max}} \Nin\} \\
		&\cup \{ k_{\text{max}} \Nin+1 ,\dots, \Nout\} 
		\end{aligned}\\
		&=: \mathcal{I}_1 \cup \dots \cup \mathcal{I}_{k_{\text{max}}+1}.
	\end{align}
Furthermore, define the operators \(P_k\), corresponding to the columns of \(G\) contained in \(\mathcal{I}_k\) (i.e. the \(k\)-th block) as
\begin{equation}
	P_k := \sum_{i=(k-1) \Nin+1}^{k \Nin} \frac{1}{\tilde{n}!} \left(\sum_{j=1}^{\Nin} G_{ij}^* a_j^{\dagger} \right)^{\tilde{n}} \dyad{0} \left(\sum_{k=1}^{\Nin} G_{ik} a_k \right)^{\tilde{n}}
\end{equation}
for \(k=1, \dots ,  k_{\text{max}}\), and 
\begin{equation}
	P_{k_{\text{max}}+1} = \sum_{i= k_{\text{max}}\Nin+1}^{\Nout} \frac{1}{\tilde{n}!} \left(\sum_{j=1}^{\Nin} G_{ij}^* a_j^{\dagger} \right)^{\tilde{n}} \dyad{0} \left(\sum_{k=1}^{\Nin} G_{ik} a_k \right)^{\tilde{n}}
\end{equation}
for the remaining terms. If the output modes are multiple of the input modes \(P_{k_{\text{max}}+1}\) will be empty.

Then, for any \(\rho^{(\tilde{n})}\)
\begin{equation}
	\begin{aligned}
		&\Tr \left[  \sum_{i=1}^{\Nout} \frac{\left(b^{\dagger}_i\right)^{\tilde{n}}}{\tilde{n}!} \dyad{0} \left(b_i\right)^{\tilde{n}} \left(\rho^{(\tilde{n})} \otimes \dyad{0} \right) \right] \\ 
		&= \sum_{k=1}^{k_{\text{max}}+1} \Tr \left[ P_k \rho^{(\tilde{n})} \right],
	\end{aligned}
\end{equation}
and hence,
\begin{equation}
	\begin{aligned}
		&\max_{\rho^{(n)}} \Tr \left[  \sum_{i=1}^{\Nout} \frac{\left(b^{\dagger}_i\right)^{\tilde{n}}}{\tilde{n}!} \dyad{0} \left(b_i\right)^{\tilde{n}} \left(\rho^{(\tilde{n})} \otimes \dyad{0} \right)\right] \\ 
		&\leq \sum_{k=1}^{k_{\text{max}}+1} \max_{\rho^{(n)}} \Tr \left[ P_k \rho^{(\tilde{n})} \right],
	\end{aligned}
\end{equation}

Now, we will make use of our partitioning in \(G\) as well. Again, note that due to \cref{eq:trace b using G only} and the particular partitioning \(\mathcal{I}\) each \(P_k\) only depends on the \(k\)-th \(\Nin \times \Nin\) block \(\Lambda_k\) of \(G\), i.e.
\begin{equation}
	G  = \begin{pmatrix}
		\Lambda_1 \\
		\hline
		\vdots \\
		\hline \\
		\Lambda_{k_{\text{max}}} \\
		\hline\\
		\Lambda_{k_{\text{max}}+1}
	\end{pmatrix},
\end{equation}
where the last block \(\Lambda_{\kmax}\) has size \((\Nout - \kmax\Nin) \times \Nin\).

One can view each \(\Lambda_k\) as a separate mode transformation for each block. However, each \(\Lambda_k\) is not unitary, still, let us perform a SVD on each \(\Lambda_k\) separately.

Therefore, for each \(k= 1, \dots, \kmax \) it holds \(\Lambda_k = U_k \Sigma_k V_k^{\dagger}\), where both \(U_k\) and \(V_k\) are now unitary matrices of dimension \(\Nin \times \Nin \) and \(\Sigma_k\) contains the singular values on the diagonal, i.e. \(\Sigma_k = \diag(\sigma_1, \dots \sigma_{\Nin})\). We will consider \(\Lambda_{\kmax}\) separately afterwards.

We will proceed with an arbitrary \(k \in \{1,\dots \kmax\}\), and note that one can absorb \(V_k^{\dagger}\) into the definition of the incoming signal modes via \(\vec{c} = V_k^{\dagger} \vec{a} \).

Hence, while stating the dependence on the modes explicitly in square brackets, it holds
\begin{equation}
	\Tr \left[ P_k[\Lambda_k \vec{a} ] \rho^{(\tilde{n})}[\vec{a}] \right] = \Tr \left[ P_k[U_k \Sigma_k\vec{c} ] \rho^{(\tilde{n})}[\vec{c}] \right].
\end{equation}

The linear mode transformation \(U_k\) induces a unitary transformation \(\Omega_{U}\) on the Hilbert space. Therefore, one can equivalently write
\begin{equation}
	\begin{aligned}
		&\Tr \left[ P_k[U_k \Sigma_k\vec{c} ] \rho^{(\tilde{n})}[\vec{c}] \right] = \Tr \left[\Omega_U  P_k[\Sigma_k\vec{c} ] \; \Omega_U^{\dagger} \rho^{(\tilde{n})}[\vec{c}]  \right] \\
		&=\Tr \left[ P_k[\Sigma_k\vec{c} ] \; \Omega_U^{\dagger} \rho^{(\tilde{n})}[\vec{c}] \Omega_U \right].
	\end{aligned}
\end{equation}
Since this can equivalently be viewed as a unitary acting on \(\rho\), the maximization can be recast as
\begin{equation}
	\max_{\rho^{(n)}} \Tr \left[ P_k \rho^{(\tilde{n})}\right] = \max_{\tau^{(n)}}
	\Tr \left[ \bar{P}_k \tau^{(\tilde{n})}\right], 
\end{equation}
where for simplicity \(\bar{P}_k := P_k[\Sigma_k \vec{c}]\) is the POVM element before the mode transformation by \(U_k\) and \(\tau^{(\tilde{n})} =  \Omega_U^{\dagger} \rho^{(\tilde{n})}[\vec{c}] \Omega_U \). 

Now, for \(k= \kmax +1\) (or equivalently \(P_{k_{\text{max}}+1}\)) the same procedure can be applied after we extend both the unitary \(U_{k_{\text{max}}+1}\) and \(\Sigma_{k_{\text{max}}+1}\) appropriately. If we extend \(U_{k_{\text{max}}+1}\) with a block of an identity of dimension \( l=  \Nout - k_{\text{max}}\Nin\) and call the resulting matrix \(\bar{U}_{k_{\text{max}}+1}\) it has the form
\begin{equation}
	\bar{U}_{k_{\text{max}}+1} = U_{k_{\text{max}}+1} \oplus \1_{l \times l}.
\end{equation}
We need to extend \(\Sigma_{k_{\text{max}}+1}\) with zeros to reach
\begin{equation}
	\bar{\Sigma}_{k_{\text{max}}+1} = \begin{pmatrix}
		\Sigma_{k_{\text{max}}+1} \\
		\hline
		0_{l \times \Nin}
	\end{pmatrix}
\end{equation}
Then, it holds
\begin{equation}
	\bar{U}_{k_{\text{max}}+1} \bar{\Sigma}_{k_{\text{max}}+1} V_{k_{\text{max}}+1}^{\dagger} = \begin{pmatrix}
		\Lambda_{k_{\text{max}}+1} \\
		\hline
		0_{l \times \Nin}
	\end{pmatrix},
\end{equation}
and the resulting mode transformation is still equivalent. We can now view \(\bar{\Sigma}_{k_{\text{max}}+1}\) as a square matrix containing singular values on the diagonal out of which some are equal to zero. Let us note that the unitary acting on the Hilbert space, \(\Omega_{\bar{U}}\), generated by \(\bar{U}_{k_{\text{max}}+1}\) acts as \( \Omega_{\bar{U}} = \Omega_{U} \otimes \1 \).
Therefore, we can still apply the same procedure as for any other \(k = 1 \dots k_{\text{max}}\).

In summary, we can write each \(\bar{P}_k\) as
\begin{equation}
	\bar{P}_k = \sum_{i=(k-1) \Nin+1}^{k \Nin} \sigma_i^{2\tilde{n}} \frac{\left(c^{\dagger}_i\right)^{\tilde{n}}}{\tilde{n}!} \dyad{0} \left(c_i\right)^{\tilde{n}},
\end{equation}
where again the \(c_i\) are the incoming modes including the unitary \(V_k^{\dagger}\) and the \(\sigma_i\) are the singular values of the block \(\Lambda_k\) of \(G\) (including zeros for \(\kmax +1\)).

Finally, the trace \(\Tr \left[ P_k \rho^{(\tilde{n})}\right]\) is maximized by
\begin{equation}
	\begin{aligned}
		&\max_{\rho^{(n)}} \Tr \left[ P_k \rho^{(\tilde{n})}\right]
		= \max_{\tau^{(n)}} \Tr \left[ \bar{P}_k \tau^{(\tilde{n})}\right] \\ 
		= &\max_{1\leq i \leq \Nin } \sigma_i^{2n} =: \lambda_k^{2n},
	\end{aligned}
\end{equation}
where for convenience we defined the singular value achieving the maximum as \(\lambda_k\). Thus, in total we find
\begin{equation}
	\min_{\rho^{(n)}} \Tr \left[ M_{\tilde{n}} \rho^{(\tilde{n})}\right] \geq 1 - \sum_{k=1}^{k_{\text{max}}+1} \lambda_k^{2n}.
\end{equation}
None of the steps presented here made any use of the particular partition \(\mathcal{I}\), we simply chose the trivial one for convenience. Moreover, all of the steps from above are still valid for any arbitrary partition \(\mathcal{I}\). Hence, we can also maximize over all possible partitions, which results in the theorem statement.
\end{proof}

\begin{lem}[Detector decomposition (restated)]
	Let  \(U \in \C^{2\Nout \times 2\Nout}\) be a unitary matrix. Furthermore, label the first \(\Nin\) columns of \(U\) as \(G\). Then, there exist \(V_{d}, V_l \in \C^{\Nout \times \Nin} \) and \(W \in \C^{2\Nin \times \Nin}\) semi-unitaries (\(W^{\dagger} W = \1\)), such that \(G\) can be written as
	\begin{equation}
		G = \left( V_{d} \oplus  V_{l} \right) W.
	\end{equation}
	Additionally, there exists a unitary matrix \(\tilde{U}\) with equal first \(\Nin\) columns written as
	\begin{equation}
		\tilde{U}  = \begin{pmatrix}
			T_d & 0 \\
			0 & T_l
		\end{pmatrix} \cdot
		\left(
		\begin{matrix}
			W_d \oplus \frac{\1}{\sqrt{2}} \hfill \\
			W_l \oplus \frac{\1}{\sqrt{2}} 
		\end{matrix}
		\, \middle\vert \,
		W'
		\right) := V_{\text{tot}} \cdot W_{\text{tot}},
	\end{equation}
	where \(W'\) is any matrix completing \(W_{\text{tot}}\) to a unitary, and both \(V_{\text{tot}} \in \C^{2\Nout \times 2\Nout} \) and \( W_{\text{tot}} \in  \C^{2\Nout \times 2\Nout}\) are unitary matrices.
\end{lem}
\begin{proof}
	At first, let us rewrite \(G\) as 
	\begin{equation}
		G = \begin{pmatrix}
			\vec{b}_1 \dots \vec{b}_{\Nin} \\
			\hline
			\vec{c}_1 \dots \vec{c}_{\Nin}
		\end{pmatrix},
	\end{equation}
	where \(\vec{b}_i , \vec{c}_i  \in \C^{\Nout}\).
	Define the vectors \(\vec{v}_i\) and \(\vec{w}_i\) for \(1\dots \Nin\)
	\begin{align}
		\beta_i := \norm{\vec{b}_i}, \quad \gamma_i := \norm{\vec{c}_i}, \\
		\vec{v}_i := \frac{\vec{b}_i}{\beta_i}, \quad \vec{w}_i := \frac{\vec{c}_i}{\gamma_i},
	\end{align}
	and using those vectors, define the matrices
	\begin{align}
		V_d':= \left( \begin{matrix}
			| \\
			\vec{v}_1 \\
			|
		\end{matrix} 
		\middle \vert \dots \middle \vert
		\begin{matrix}
			| \\ 
			\vec{v}_{\Nin}  \\
			|
		\end{matrix}
		\right) \text{ and } 
		V_l':= \left( \begin{matrix}
			| \\
			\vec{w}_1 \\
			|
		\end{matrix} 
		\middle \vert \dots \middle \vert
		\begin{matrix}
			| \\ 
			\vec{w}_{\Nin}  \\
			|
		\end{matrix}
		\right).
	\end{align}
	Then, we can rewrite \(G\) as
	\begin{equation}
		G = \begin{pmatrix}
			V'_d & 0 \\
			0 & V_l'
		\end{pmatrix}
		\begin{pmatrix}
			\diag(\vec{\beta}) \\
			\diag(\vec{\gamma})
		\end{pmatrix},
	\end{equation}
	where \(\vec{\beta}\) and  \(\vec{\gamma}\) are vectors having \(\beta_i\) and \(\gamma_i\) as their components. These matrices are no semi-unitaries yet, hence take the polar decomposition of \(V_d'\) and \(V_l'\) yielding
	\begin{align}
		V_d' = V_d  P_d, \quad \text{and} \quad V_l' = V_l  P_l.
	\end{align}
	To simplify the notation, let us define \(W_d := P_d \diag(\vec{\beta})\) and \(W_l := P_l \diag(\vec{\gamma})\). This allows us to write \(G\) as
	\begin{equation}
		G = \begin{pmatrix}
			V_d & 0 \\
			0 & V_l
		\end{pmatrix}
		\begin{pmatrix}
			W_d \\
			W_l			
		\end{pmatrix} =:\begin{pmatrix}
			V_d & 0 \\
			0 & V_l
		\end{pmatrix} \cdot W
	\end{equation}
	By definition \(V_d\) and \(V_l\) are semi-unitaries. Also, \(W \) is semi-unitary since
	\begin{align}
		\begin{aligned}
			W^{\dagger} W = &\diag(\vec{\beta}) P_d^{\dagger} P_d \diag(\vec{\beta}) \\ &+\diag(\vec{\gamma}) P_l^{\dagger} P_l \diag(\vec{\gamma}) \\
			=& \diag(\vec{\beta}) V_d^{\dagger} P_d \diag(\vec{\beta}) \\ &+\diag(\vec{\gamma}) V_l^{\dagger} P_l \diag(\vec{\gamma}) \\
			= \; &G^{\dagger} G = \1,
		\end{aligned} 
	\end{align}
	where we used \(P_{d/l}^{\dagger} P_{d/l} = V_{d/l}^{\dagger} V_{d/l} \), by the polar decomposition. Hence, we found the desired decomposition of \(G\).
	
	Next we want to extend this semi-unitary to a full unitary, generating the first \(\Nin\) columns (or equivalently \(G\)).
	Using the QR-decomposition, we can write
	\begin{align}
		V_d = Q_d R \quad \text{ and } \quad V_l = Q_l R,
	\end{align}
	where both \( Q_d \) and \( Q_l \) are unitary \(\Nout \times \Nout\) matrices. The matrix \(R\) is composed of an identity and a block of zeros \(O\) of size \({\left(\Nout-\Nin\right) \times \Nin}\), i.e.
	\begin{equation}
		R = \begin{pmatrix}
			\1_{\Nin} \\ 
			\hline
			O
		\end{pmatrix}.
	\end{equation} 
	With that define the matrix \(V_{\text{tot}}\) as
	\begin{equation}
		V_{\text{tot}} = \begin{pmatrix}
			Q_d & 0 \\
			0 & Q_l
		\end{pmatrix},
	\end{equation}
	which is a unitary matrix because both \(Q_d\) and \(Q_l\) are unitary. Now, we first extend \(W\) to
	\begin{align}
		W_{\text{ext}} := \begin{pmatrix}
			W_d \oplus \frac{\1_{\Nout-\Nin}}{\sqrt{2}} \\
			W_l \oplus \frac{\1_{\Nout-\Nin}}{\sqrt{2}}
		\end{pmatrix}
	\end{align}
	and then take another QR decomposition of \(W_{\text{ext}}\) such that
	\begin{align}
		W_{\text{ext}} = W_{\text{tot}} \cdot \begin{pmatrix}
			\1_{\Nout} \\
			\hline
			O_{\Nout}
		\end{pmatrix},
	\end{align}
	where \(O_{\Nout}\) is a matrix containing only zeros of size \(\Nout \times \Nout \). Finally, the unitary \(\tilde{U} := V_{\text{tot}} \cdot W_{\text{tot}}\) results in the same \(G\), since
	\begin{align}
		\begin{aligned}
			&V_{\text{tot}} W_{\text{tot}} \cdot \begin{pmatrix}
				\1_{\Nin} \\
				0
			\end{pmatrix} = \begin{pmatrix}
				Q_d & 0 \\
				0 & Q_l
			\end{pmatrix}\cdot
			\begin{pmatrix}
				W_d \\
				0 \\
				\hline
				W_l \\
				0
			\end{pmatrix}\\
			= &\begin{pmatrix}
				Q_d \begin{pmatrix}
					W_d \\
					0
				\end{pmatrix} \\
				Q_l \begin{pmatrix}
					W_l \\
					0
				\end{pmatrix}
			\end{pmatrix} = 
			\begin{pmatrix}
				V_d \cdot W_d \\
				V_l \cdot W_l
			\end{pmatrix} 
			= G.
		\end{aligned}
	\end{align}
\end{proof}

\begin{thrm}[Subspace estimation for linear optics detection setup with loss (restated)]
	Let  \(U \in \C^{2\Nout \times 2\Nout}\) be a unitary matrix describing the mode transformation of a passive linear optical setup with losses as in \cref{eq:Linear mode TF}. Let \(V_d \in \C^{\Nout \times \Nin}\) be the semi-unitary resulting from applying \cref{Lem:Detector decomposition} to \(U\). By using the POVM element 
	\begin{equation}
		M^{\mathrm{mult}}(V_d) := \1 - \dyad{0} - \sum_{i=1}^{\Nout} F_{\text{sc}}^{i}.
	\end{equation}
	of \cref{eq:POVM flag-state lossy}, defined in terms of \(V_d\), and assuming threshold detectors, the \( \tilde{n}\leq N_B \)-subspace can be bounded by
	\begin{equation}
		p_L(\tilde{n} \leq N_B|x) = 1 - \frac{m_{\mathrm{mult}|x}}{c_{\geq N_B +1}},
	\end{equation}
	where \(c_{\geq N_B +1} := \min_{{\tilde{n}}\geq N_B +1} c_{\tilde{n}} \). The constants \(c_{\tilde{n}}\) depend on the partitions \(\mathcal{I} = \bigcup_{k=1}^{k_{\text{max}} +1 } \mathcal{I}_k\), where \( k_{\text{max}} = \lfloor \frac{\Nout}{\Nin}\rfloor\). These partitions are of size \(|\mathcal{I}_k| =\Nin\) for \(k = 1,\dots, \kmax\) and \(|\mathcal{I}_{\kmax +1 }| = \Nout -\kmax \Nin\). The constants \(c_{\tilde{n}}\) are defined as
	\begin{equation}
		\begin{aligned}
			c_{\tilde{n}} &= \max_{\mathcal{I} = \cup \mathcal{I}_k} \left(1 - \sum_{k=1}^{k_{\text{max}}+1} \lambda^{2n}\left(\mathcal{I}_k\right) \right), \quad n\geq 0, \\
			c_0 &= 0,
		\end{aligned}
	\end{equation}
	where \(\lambda\left(\mathcal{I}_k\right)\) is the maximum singular value of the \(\Nin \times \Nin \) block of \(V_d\), containing all rows labelled by \(\mathcal{I}_k\).
\end{thrm}
\begin{proof}
	At first, let us represent \(U\) using \cref{Lem:Detector decomposition} as
	\begin{equation}
		\tilde{U} = \begin{pmatrix}
			T_d & 0 \\
			0 & T_l
		\end{pmatrix}\cdot \left(
		\begin{matrix}
			W_d \oplus \frac{\1}{\sqrt{2}} \hfill \\
			W_l \oplus \frac{\1}{\sqrt{2}} 
		\end{matrix}
		\, \middle\vert \,
		W'
		\right)  =:  V_{\text{tot}} \cdot W_{\text{tot}}.
	\end{equation} 
	Next, let us define the mode operators in the following way, let \(\vec{a}\) be the incoming modes (including the \(\Nin\) signal modes), \(\vec{b}\) the transformed modes after the loss, i.e. after applying \(W_{\text{tot}}\), and \(\vec{c}\) be the modes after also \(V_{\text{tot}}\) has been applied. Thus, it holds \(\vec{c} = \tilde{U} \vec{a} \) and \( \tilde{U} \vec{a}\vert_{\Nin} = U \vec{a}\vert_{\Nin}\) for the incoming signal modes only.
	Moreover, let us label the system of the incoming \(\Nin\) signal modes by \(\text{Sig}\), the system of vacuum modes \(\Nin + 1, \dots \Nout\) by \(\text{Env}_1\) and the remaining incoming modes by \(\text{Env}_2\).
	
	Next, consider the POVM element \(M^{\text{full}}\) acting on the full system of output modes \(\vec{c}\), which has the form
	\begin{equation}
		M^{\text{full}} = M^{\text{det}} \otimes \1_{\text{Env}_2},
	\end{equation}
	where \(M^{\text{det}}\) acts on the \(\Nout\) detected modes and the identity on the \(\Nout\) loss modes. Finally, let for any unitary transformation \(V\) acting on the modes, let \(\Omega_V\) be the corresponding transformation on the Hilbert space. Then, for any state \(\rho_{\text{sig}}\) defined on the \(\Nin\) signal input modes in the \(\text{Sig}\) system, we find
	\begin{align}
		&\Tr\left[ M^{\text{full}} \Omega_{\tilde{U}} \left(\rho_{\text{sig}} \otimes \dyad{0}_{\text{Env}_1 \text{Env}_2} \right) \Omega_{\tilde{U}}^{\dagger} \right] \\ 
		=& \Tr\left[ \left(\Omega_{T_d}^{\dagger} M^{\text{det}} \Omega_{T_d} \otimes \1_{\text{Env}_2} \right) \Omega_{W} \right. \nonumber \\
		&\hspace{30pt} \cdot \left. \left(\rho_{\text{sig}} \otimes \dyad{0}_{\text{Env}_1 \text{Env}_2} \right) \Omega_{W}^{\dagger} \right] \\ 
		=& \Tr\left[ \Omega_{T_d}^{\dagger} M^{\text{det}} \Omega_{T_d} \right. \nonumber \\
		&\hspace{20pt} \cdot \left. \Tr_{\text{Env}_2} \left[\Omega_{W} \left(\rho_{\text{sig}} \otimes \dyad{0}_{\text{Env}_1 \text{Env}_2} \right) \Omega_{W}^{\dagger} \right] \right] \\
		=& \Tr\left[ \Omega_{T_d}^{\dagger} M^{\text{det}} \Omega_{T_d} \left(\Psi\left(\rho_{\text{in}} \right) \otimes \dyad{0}_{\text{Env}_1}\right) \right],
	\end{align}
	where we used that \(W_{\text{tot}}\) and thus \(\Omega_{W}\) acts as the identity on the incoming vacuum modes in system \(\text{Env}_1\). Furthermore, we defined the CPTP-map \(\Psi\) as
	\begin{equation}
		\Psi(\rho_{\text{sig}}) := \Tr_{\text{Env}_1 \text{Env}_2}\left[ \Omega_{W} \left(\rho_{\text{sig}} \otimes \dyad{0}_{\text{Env}_1 \text{Env}_2} \right) \Omega_{W}^{\dagger}  \right],
	\end{equation}
	such that 
	\begin{equation}
		\begin{aligned}
			&\Tr_{\text{Env}_2} \left[\Omega_{W} \left(\rho_{\text{sig}} \otimes \dyad{0}_{\text{Env}_1 \text{Env}_2} \right) \Omega_{W}^{\dagger} \right] \\
			= &\Psi(\rho_{\text{sig}}) \otimes \dyad{0}_{\text{Env}_1}.
		\end{aligned}
	\end{equation}
	The transformed part \(\Omega_{T_d}^{\dagger} M^{\text{det}} \Omega_{T_d}\) of \(M^{\text{full}}\) acting on the detected modes only is given by
	\begin{equation}
		\Omega_{T_d}^{\dagger} M^{\text{det}} \Omega_{T_d} = M^{\mathrm{mult}}(V_d) = \1 - \dyad{0} - \sum_{i=1}^{\Nout} F_{\text{sc}}^{i},
	\end{equation}
	where each \(F_{\text{sc}}^{i}\) is
	\begin{equation}
		F_{\text{sc}}^{i} = \sum_{n=1}^{\infty} \frac{\left(c^{\dagger}_i\right)^n}{n!} \dyad{0} \left(c_i\right)^n,
	\end{equation}
	using our convention for labelling the modes from before. The subset of the output modes \(\vec{c}_d\) corresponding to the detected modes is given by \(\vec{c}_d = T_d \vec{b}_d\). Thus, we still implicitly assume the modes \(c_i\) as a function of the modes \(b_j\). Finally, we find for the constants \(c_{\tilde{n}}\), where \(\tilde{n}\) is now to be understood as \(\tilde{n}\) photons after \(W_{\text{tot}}\) was applied,
	\begin{align}
		c_{\tilde{n}} &= \min_{\substack{\rho_{\text{sig}} \\ \text{s.t. } \Psi\left(\rho_{\text{in}} \right) = \tau^{(\tilde{n})}}} \Tr\left[  M_{\tilde{n}}(V_d) \left(\tau^{(\tilde{n})} \otimes \dyad{0}_{\text{Env}_1}\right) \right] \\ 
		&\geq \min_{ \tau^{(\tilde{n})}} \Tr\left[  M_{\tilde{n}}(V_d) \left(\tau^{(\tilde{n})} \otimes \dyad{0}_{\text{Env}_1}\right) \right].
	\end{align}
	Now, can apply \cref{Thrm:Lower bound flag state perfect} on \(M_n(V_d)\), while optimizing \(\tau^{(\tilde{n})}\).
	Additionally, the quantum channel \(\Psi\) will also be included in the squashing map \(\Lambda\) and thus be given to Eve.
\end{proof}

\begin{cor}[Lower bound including dark counts (restated)]
	Let \(G \in \C^{\Nout \times \Nin}\) be the submatrix describing a detection setup consisting only of passive linear optical elements. Furthermore, let \(\mathcal{P}\) be the stochastic matrix implementing the post processing corresponding to dark counts as in \cref{eq:PP dark counts}. Then, the lower bounds on the subspaces of the previous \cref{Thrm:Lower bound flag state perfect} and \cref{Thrm:Lower bound flag state lossy} are still valid using observations including dark counts and the constants \(c_{\tilde{n}}\) from the setups without dark counts.
\end{cor}
\begin{proof}
	Assume we define \(M^{\text{dark}}\) as in \cref{eq:POVM flag-state perfect}, but using the POVM elements \(\{\tilde{F}_j\}\) including the dark count post-processing and \(M\) just as before. With this construction, the \(\tilde{n}\)-photon component of \(M^{\text{dark}}\) is a linear combination of \(\tilde{n}\)-photon components of the POVM elemnts \(\{F_i\}\) before the post-processing. Thus, it follows
	\begin{equation}
		\begin{aligned}
			M_{\tilde{n}}^{\text{dark}} &= \Pi_{\tilde{n}} - \sum_{k,i=0}^{N_{\text{out}}} \mathcal{P}_{ki} \frac{\left(b^{\dagger}_i\right)^{\tilde{n}}}{\tilde{n}!} \dyad{0} \left(b_i\right)^{\tilde{n}} \\
			&=  \Pi_{\tilde{n}} - \sum_{i=0}^{N_{\text{out}}} \left( \sum_{k=0}^{N_{\text{out}}} \mathcal{P}_{ki} \right) \frac{\left(b^{\dagger}_i\right)^{\tilde{n}}}{\tilde{n}!} \dyad{0} \left(b_i\right)^{\tilde{n}} \\
			&=  \Pi_{\tilde{n}} - \sum_{i=0}^{N_{\text{out}}} p_i \frac{\left(b^{\dagger}_i\right)^{\tilde{n}}}{\tilde{n}!} \dyad{0} \left(b_i\right)^{\tilde{n}} \geq M_{\tilde{n}},
		\end{aligned}
	\end{equation}
	where we used \(\sum_{i} \mathcal{P}_{ki} = p_i \leq 1 \; \forall j\). Hence, we can conclude
	\begin{align}
		&\Tr \left[ M_{\tilde{n}}^{\text{dark}} \rho^{(\tilde{n})}\right] \geq  \Tr \left[ M_{\tilde{n}}^{\mathrm{mult}} \rho^{(\tilde{n})}\right] \geq c_{\tilde{n}}.
	\end{align}
	Since this holds for all \(\tilde{n} \in \N\) it follows for the observations of the POVM element with and without dark counts \(m_{\text{obs}}^{\text{dark}} \geq m_{\text{obs}}\). Hence, we can write
	\begin{equation}
		p_L(\tilde{n} \leq N_B|x) = 1 - \frac{m_{\mathrm{mult}|x}^{\text{dark}}}{c_{\geq N_B +1}^{\text{dark}}} \geq  1 - \frac{m_{\mathrm{mult}|x}^{\text{dark}}}{c_{\geq N_B +1}},
	\end{equation}
	which concludes the proof.
\end{proof}

\section{Detector Characterization for Flag-State Squasher}\label{App:Detector Characterization for Flag-State Squasher}
In this section, we answer the question of how one could determine the detector matrix \(G\) experimentally. Since there will always be some loss involved in any experimental setup, we only discuss the more general lossy case here.

At first, consider a coherent state with complex amplitude \(\vec{\alpha}\) in the input modes entering the detection setup and it is transformed according to \(G\) via 
\begin{equation}
	\beta_i = \sum_{j=1}^{N_{\text{in}}} G_{ij} \alpha_j,
\end{equation}
assuming there is no stray light entering the setup. For threshold detectors the probability of observing a click in detector \(i\) corresponding to mode \(b_i\) originating from a coherent laser pulse is given by
\begin{equation}\label{eq:Prob detector i}
	p_i = 1- e^{-|\beta_i|^2} = 1- e^{-|\sum_{j=1}^{N_{\text{in}}} G_{ij} \alpha_j|^2}.
\end{equation}
Thus, with the knowledge of the incoming complex amplitude \(\vec{\alpha}\) of a coherent state and the probability of a detection, one can solve for \(G\), provided enough test states are sent. We do not have access to the exact probabilities, but we can use normalized counts in the detectors as an approximation for the detection probability. In the limit of many pulses, this will converge to the probability \(p_i\) by the law of large numbers. Furthermore, if one uses a bright laser or assumes low dark count rates, one can neglect recorded counts due to dark counts.

The only question that remains is how many different states need to be sent in order to characterize the detector matrix \(G\) sufficiently. Any detector matrix \(G\) can be split horizontally into a detection \(G_d\) and a loss part \(G_l\), such that
\begin{equation}
	G = \begin{pmatrix}
		G_d \\
		\hline
		G_l
	\end{pmatrix}.
\end{equation}
Then, let us apply the singular value decomposition on \(G_l\) to find
\begin{equation}
	G_l = U_l \Sigma_l R_l,
\end{equation}
where \(\Sigma_l\) contains the singular values \(\vec{\sigma}\) on the diagonal and in this case has the special form
\begin{equation}
	\Sigma_l = \begin{pmatrix}
		\diag(\vec{\sigma}) \\
		\hline
		0_{\left(\Nout -\Nin \right) \times \Nin }
	\end{pmatrix}.
\end{equation}
We can combine \(R_l\) and \(\Sigma_l\), thus let us define \(Q_l := \diag(\vec{\sigma}) R_l\). Therefore, the detector matrix \(G\) can equivalently be written as
\begin{equation}\label{eq:G up to unitary in loss}
	G = \begin{pmatrix}
		G_d \\
		\hline
		G_l
	\end{pmatrix}
	= \begin{pmatrix}
		\1_d & 0 \\
		0 & U_l
	\end{pmatrix} \cdot 
	\begin{pmatrix}
		G_d  \\
		\hline
		Q_l \\
		0
	\end{pmatrix}
	=: \begin{pmatrix}
		\1_d & 0 \\
		0 & U_l
	\end{pmatrix} \cdot G'.
\end{equation}
Hence, after redefining only the loss modes as the ones after the unitary \(U_l\) has been applied, it suffices to characterize \(G'\). This reduces the free parameters by quite a bit, since now only \(2\Nin\cdot\left(\Nout+\Nin\right)\) free real parameters need to be found.

In fact this can be further restricted, since we only need to characterize each row of \(G'\) up to a global phase and moreover we can use \(\left(G'\right)^{\dagger} G' = \1_{\Nin}\). This already gives us \(\Nout+\Nin+\Nin^2\) constraints. Hence, we need to determine
\begin{equation}
	n_{\text{free}} = \Nin\left(2\Nout +\Nin -1 \right) -\Nout
\end{equation}
free real parameters during testing. Due to \cref{eq:Prob detector i}, each states generates \(\Nout\) constraints from observations. Therefore, one needs to probe a detection setup with
\begin{equation}
	n_{\text{states}} = \lceil \frac{\Nin}{\Nout} \left(2\Nout +\Nin -1 \right) - 1 \rceil
\end{equation}
distinct states to find \(G'\) up to a global phase in each row. Then, \cref{Lem:Detector decomposition} allows us to find a decomposition of \(G'\). If we combine the resulting decomposition of \(G'\) with \cref{eq:G up to unitary in loss} we reach a valid decomposition of \(G\), again up to global phases in each row. Therefore, \(n_{\text{states}}\)-many states are also required for characterizing \(G\).

We can generalise this idea to a procedure to characterize a passive linear optical detection setup with losses and threshold detectors with the steps below.
\begin{enumerate}
	\item Determine \(\vec{\alpha}_1, \dots \vec{\alpha}_{n_{\text{states}}}\) different complex amplitudes.
	\item Shine a coherent laser pulse with complex amplitude \(\vec{\alpha}_j\) in the input modes into the detection setup and record the counts in the output ports corresponding to modes \(b_i\), \(i=1,\dots N_{\text{out}}\). 
	\item Convert the counts to normalized counts, which approximate the detection probability and connect to \(G\) by
	\begin{equation*}
		p_i = 1- e^{-|\beta_i|^2} = 1- e^{-|\sum_{j=1}^{N_{\text{in}}} G_{ij} \alpha_j|^2}.
	\end{equation*}
	\item Repeat steps 2. and 3. until sufficient statistics for a good approximation of the probabilities \(p_i\) is reached.
	\item Repeat 2. -- 4. for \(j=1 \dots n_{\text{states}}\) such that one has sufficient data to solve for \(G\).
	\item Then, use \cref{Lem:Detector decomposition} to find \(V_d\) and \cref{Thrm:Lower bound flag state lossy} to calculate the lower bound minimizing over \(G\) within its uncertainties. Finally, use \(V_d\) to construct Bob's POVM elements.
\end{enumerate}

One final remark needs to be made here, since \(V_d\) will be used to construct Bob's POVM and therefore for calculating the key rate, this allows us to incorporate any passive linear optical detection setup into the numerical frame work of \cite{Winick2018Quantum}.

\section{Kraus Operators BB84 protocol}
Again, we are using the framework of \cite{Winick2018Quantum} together with the simplifications of \cite[App. A]{Lin2019Phys.Rev.X}. Furthermore, as mentioned in the main text, we use the squashing model from \cite{Gittsovich2014Phys.Rev.A, Beaudry2008Phys.Rev.Lett.} and recover Bob's qubit POVM elements, which are
\begin{equation}
	\begin{aligned}
		F^B_{(Z,0)} &= p_z \begin{pmatrix} 0 & 0 &0 \\ 0& 1 & 0\\ 0&0 & 0 \end{pmatrix}, \;
		F^B_{(Z,1)} = p_z \begin{pmatrix} 0 & 0& 0\\ 0& 0 & 0\\ 0&0 & 1 \end{pmatrix}, \\
		F^B_{(X,0)} &= \frac{p_x}{2} \begin{pmatrix} 0&0 & 0 \\  0&1 & 1 \\ 0& 1 & 1 \end{pmatrix},
		\;
		F^B_{(X,1)} = \frac{p_x}{2} \begin{pmatrix} 0&0 &0   \\  0&1 & -1  \\ 0 & -1 & 1  \end{pmatrix}, \\
		F^B_{\bot} &=\begin{pmatrix} 1 & 0 & 0 \\ 0 & 0 & 0 \\ 0 & 0 & 0 \end{pmatrix}.
	\end{aligned}
\end{equation}
For Alice's side, the POVM elements are determined by the source replacement scheme \cite{Bennett1992Phys.Rev.Lett., Ferenczi2012Phys.Rev.A} and given by 
\begin{equation}
	\begin{aligned}
		F^A_{(Z,0)} &= \ketbra{0}, F^A_{(Z,1)} = \ketbra{1}, \\
		F^A_{(X,0)} &= \ketbra{2}, F^A_{(X,1)} = \ketbra{3}.
	\end{aligned}
\end{equation}
The resulting Kraus operators are
\begin{equation}
	\begin{aligned}
		K_Z &= \left[
		\begin{pmatrix} 1 \\ 0 \end{pmatrix}
		_R  \otimes  \begin{pmatrix} 1 &  & &  \\ & 0 & & \\
			& & 0 & \\
			& & & 0 \end{pmatrix}_A +
		\begin{pmatrix} 0 \\ 1 \end{pmatrix}
		_R \otimes  \begin{pmatrix} 0 &  & &  \\ & 1 & & \\
			& & 0 & \\
			& & & 0 \end{pmatrix}_A \right] \\
		&\otimes \sqrt{p_z}
		\begin{pmatrix} 0 & & \\ & 1 & \\& & 1 \end{pmatrix}
		_B \otimes
		\begin{pmatrix} 1 \\ 0 \end{pmatrix}
		_{C}, \\
		K_X &= \left[
		\begin{pmatrix} 1 \\ 0 \end{pmatrix}
		_R  \otimes  \begin{pmatrix} 0 &  & &  \\ & 0 & & \\
			& & 1 & \\
			& & & 0 \end{pmatrix}_A +
		\begin{pmatrix} 0 \\ 1 \end{pmatrix}
		_R \otimes  \begin{pmatrix} 0 &  & &  \\ & 0 & & \\
			& & 0 & \\
			& & & 1 \end{pmatrix}_A \right] \\
		&\otimes \sqrt{p_x}
		\begin{pmatrix} 0 & & \\ & 1 & \\& & 1 \end{pmatrix}
		_B  \otimes
		\begin{pmatrix} 0 \\ 1 \end{pmatrix}
		_{C},
	\end{aligned}
\end{equation}
and
\begin{equation}
	\begin{aligned}
		Z_1 &=
		\begin{pmatrix} 1 & \\ & 0 \end{pmatrix}
		\otimes \1_{\dim_A \times\dim_B \times 2}, \\
		Z_2 &=
		\begin{pmatrix} 0 & \\ & 1 \end{pmatrix}
		\otimes \1_{\dim_A \times\dim_B \times 2}.
	\end{aligned}
\end{equation}

\section{POVM Elements and Kraus operators Six-State Protocol}\label{App:POVMs and Kraus ops 6-state}
For simplicity, we only present the Kraus operators for the case \(N_B = 1\). In all cases, the original POVM elements (before squashing) can be found in \cite{Gittsovich2014Phys.Rev.A}.

After applying the flag-state squasher, the POVM elements on Bob's \(\leq N_B\)-photon subspace are
\begin{equation}
	\begin{aligned}
		\tilde{F}^{B}_{(Z,0)} &= p_z \begin{pmatrix} 0 & 0 &0 \\ 0& 1 & 0\\ 0&0 & 0 \end{pmatrix}, \;
		\tilde{F}^{B}_{(Z,1)} = p_z \begin{pmatrix} 0 & 0& 0\\ 0& 0 & 0\\ 0&0 & 1 \end{pmatrix}, \\
		\tilde{F}^{B}_{(X,0)} &= \frac{p_x}{2} \begin{pmatrix} 0&0 & 0 \\  0&1 & 1 \\ 0& 1 & 1 \end{pmatrix},
		\;
		\tilde{F}^{B}_{(X,1)} = \frac{p_x}{2} \begin{pmatrix} 0&0 &0   \\  0&1 & -1  \\ 0 & -1 & 1  \end{pmatrix}, \\
		\tilde{F}^{B}_{(Y,0)} &= \frac{p_y}{2} \begin{pmatrix} 0&0 & 0 \\  0&1 & -i \\ 0& i & 1 \end{pmatrix},
		\;
		\tilde{F}^{B}_{(Y,1)} = \frac{p_y}{2} \begin{pmatrix} 0&0 &0   \\  0&1 & i  \\ 0 & -i & 1  \end{pmatrix}, \\
		\tilde{F}^{B}_{\bot} &=\begin{pmatrix} 1 & 0 & 0 \\ 0 & 0 & 0 \\ 0 & 0 & 0 \end{pmatrix}.
	\end{aligned}
\end{equation}
These need to be padded with flags. Therefore, let \(E_i:=\diag(0,\dots,0,1,0, \dots ,0) \in \R^{10 \times 10}\), be a diagonal matrix with \(1\) at the \(i\)-th entry. Furthermore, one needs to include double (dc) and cross-clicks (cc). For \(N_B=1\), these are only part of the flags. Hence, Bob's full squashed POVM elements are
\begin{equation}
	\begin{aligned}
		F^{B }_{(Z,0)} &= \tilde{F}^{B}_{(Z,0)} \oplus E_1, \quad
		F^{B }_{(Z,1)} = \tilde{F}^{B}_{(Z,1)} \oplus E_2, \\
		F^{B}_{(X,0)} &= \tilde{F}^{B}_{(X,0)} \oplus E_3,
		\quad
		F^{B }_{(X,1)} = \tilde{F}^{B}_{(X,1)} \oplus E_4, \\
		F^{B}_{(Y,0)} &= \tilde{F}^{B}_{(Y,0)} \oplus E_5,
		\quad
		F^{B}_{(Y,1)} = \tilde{F}^{B}_{(Y,1)} \oplus E_6, \\
		F^{B}_{\text{dc},Z} &= \bar{0}_3 \oplus E_7, \quad 
		F^{B}_{\text{dc},X} = \bar{0}_3 \oplus E_8, \\
		F^{B}_{\text{dc},Y} &= \bar{0}_3 \oplus E_9, \\
		F^{B}_{\text{cc}} &= \bar{0}_3 \oplus E_{10}, \\
		F^{B}_{\bot} &= \tilde{F}^{B}_{\bot} \oplus \bar{0}_{10},
	\end{aligned}
\end{equation}
where \(\bar{0}_m\) indicates a matrix with only zeros of dimension \(m\). Regarding Alice's POVM elements, for numerical stability we apply a Schmidt decomposition after the source replacement scheme to reduce the dimensions and arrive at
\begin{equation}
	\begin{aligned}
		F^A_{(Z,0)} &= p_z \begin{pmatrix} 1 & 0\\ 0 & 0 \end{pmatrix}, F^A_{(Z,1)} = p_z \begin{pmatrix} 0 & 0\\ 0 & 1 \end{pmatrix}, \\
		F^A_{(X,0)} &= \frac{p_x}{2} \begin{pmatrix} 1 & 1\\ 1 & 1 \end{pmatrix}, F^A_{(X,1)} =  \frac{p_x}{2} \begin{pmatrix} 1 & -1\\ -1 & 1 \end{pmatrix}, \\
		F^A_{(Y,0)} &=  \frac{p_y}{2} \begin{pmatrix} 1 & i\\ -i & 1 \end{pmatrix}, F^A_{(Y,1)} =  \frac{p_y}{2} \begin{pmatrix} 1 & -i\\ i & 1 \end{pmatrix}.
	\end{aligned}
\end{equation}
The resulting Kraus operators in the framework of \cite{Winick2018Quantum} are
\begin{equation}
	\begin{aligned}
		K_Z &= \sqrt{p_z} \left[
		\begin{pmatrix} 1 \\ 0 \end{pmatrix}
		_R  \otimes \begin{pmatrix} 1 & 0\\ 0 & 0 \end{pmatrix} +
		\begin{pmatrix} 0 \\ 1 \end{pmatrix}
		_R \otimes  \begin{pmatrix} 0 & 0\\ 0 & 1 \end{pmatrix} \right] \\
		&\otimes \left[0 \oplus \sqrt{p_z} \1_2 \oplus \left(E_1 + E_2 \right) \right]_B
	 	\otimes \begin{pmatrix} 1 \\ 0 \\ 0 \end{pmatrix}_{C}, \\
		K_X &= \frac{ \sqrt{p_x}}{2} \left[
		\begin{pmatrix} 1 \\ 0 \end{pmatrix}
		_R  \otimes   \begin{pmatrix} 1 & 1\\ 1 & 1 \end{pmatrix} +
		\begin{pmatrix} 0 \\ 1 \end{pmatrix}
		_R \otimes \begin{pmatrix} 1 & -1\\ -1 & 1 \end{pmatrix} \right] \\
		&\otimes  \left[0 \oplus \sqrt{p_x} \1_2 \oplus \left(E_3 + E_4 \right) \right]_B
		\otimes
		\begin{pmatrix} 0 \\ 1 \\ 0 \end{pmatrix}
		_{C},\\
		K_Y &= \frac{\sqrt{p_y}}{2} \left[
		\begin{pmatrix} 1 \\ 0 \end{pmatrix}
		_R  \otimes  \begin{pmatrix} 1 & i\\ -i & 1 \end{pmatrix} +
		\begin{pmatrix} 0 \\ 1 \end{pmatrix}
		_R \otimes  \begin{pmatrix} 1 & -i\\ i & 1 \end{pmatrix} \right] \\
		&\otimes \left[0 \oplus \sqrt{p_y} \1_2 \oplus \left(E_5 + E_6 \right) \right]_B
		\otimes
		\begin{pmatrix} 0 \\ 0 \\ 1 \end{pmatrix}
		_{C},
	\end{aligned}
\end{equation}
and
\begin{equation}
	\begin{aligned}
		Z_1 &=
		\begin{pmatrix} 1 & \\ & 0 \end{pmatrix}
		\otimes \1_{\dim_A \times\dim_B \times 3}, \\
		Z_2 &=
		\begin{pmatrix} 0 & \\ & 1 \end{pmatrix}
		\otimes \1_{\dim_A \times\dim_B \times 3}.
	\end{aligned}
\end{equation}

\end{document}